\documentclass[galaxies,article,accept,pdftex,moreauthors]{Definitions/mdpi}

\firstpage{1} 
\pubvolume{1}
\issuenum{1}
\articlenumber{0}
\pubyear{2025}
\copyrightyear{2025}
\datereceived{ } 
\daterevised{ } 
\dateaccepted{ } 
\datepublished{ } 
\hreflink{https://doi.org/} 

\usepackage[T1]{fontenc}
\usepackage[utf8]{inputenc}
\usepackage{lmodern}
\DeclareUnicodeCharacter{2212}{-}

\usepackage{amsmath}
\usepackage{url}
\DeclareRobustCommand{\mytt}[1]{\ifmmode\text{\path{#1}}\else\path{#1}\fi}

\definecolor{valecol}{rgb}{0,0.5, 1.}

\graphicspath{{./figs/}}

\Title{The miniJPAS and J-NEP surveys: Machine learning for star-galaxy separation}

\TitleCitation{The miniJPAS and J-NEP surveys: Machine learning for star-galaxy separation}

\Author{
Ana Paula Jeakel$^{1,\ddagger}$\orcidA{},
Gabriel Vieira dos Santos$^{2,\ddagger}$\orcidB{},
Valerio Marra$^{2,3,4,*}$\orcidC{},
Rodrigo von Marttens$^{5}$\orcidD{},
Siddhartha Gurung-López$^{6,7}$,
Raul Abramo$^{8}$,
Jailson Alcaniz$^{9}$,
Narciso Benitez$^{10}$,
Silvia Bonoli$^{11, 12}$,
Javier Cenarro$^{12,13}$,
David Cristóbal-Hornillos$^{12}$,
Simone Daflon$^{9}$,
Renato Dupke$^{9}$,
Alessandro Ederoclite$^{12,13}$,
Rosa M.\ González Delgado$^{10}$,
Antonio Hernán-Caballero$^{12,13}$,
Carlos Hernández–Monteagudo$^{15}$,
Jifeng Liu$^{15}$,
Carlos López-Sanjuan$^{12,13}$,
Antonio Marín-Franch$^{12,13}$,
Claudia Mendes de Oliveira$^{16}$,
Mariano Moles$^{12}$,
Fernando Roig$^{9}$,
Laerte Sodré Jr.$^{16}$,
Keith Taylor$^{17}$,
Jesús Varela$^{12}$,
Héctor Vázquez Ramió$^{12,13}$,
José M.\ Vilchez$^{10}$,
Christopher Willmer$^{18}$,
Javier Zaragoza-Cardiel$^{12}$
}

\AuthorCitation{Jeakel, Vieira dos Santos et al.}

\address{%
$^{1}$ \quad PPGFis, Universidade Federal do Espírito Santo, 29075-910, Vitória, ES, Brazil\\
$^{2}$ \quad Departamento de Física, Universidade Federal do Espírito Santo, 29075-910, Vitória, ES, Brazil\\
$^{3}$ \quad INAF -- Osservatorio Astronomico di Trieste, via Tiepolo 11, 34131 Trieste, Italy\\
$^{4}$ \quad IFPU -- Institute for Fundamental Physics of the Universe, via Beirut 2, 34151, Trieste, Italy\\
$^{5}$ \quad Instituto de Física, Universidade Federal da Bahia, 40210-340, Salvador, BA, Brazil\\
$^6$ \quad Observatori Astron\`omic de la Universitat de Val\`encia, Ed. Instituts d’Investigaci\'o, Parc Cient\'ific. C/ Catedr\'atico Jos\'e Beltr\'an, n2, 46980 Paterna, Valencia, Spain\\
$^7$ \quad Departament d’Astronomia i Astrof\'isica, Universitat de Val\`encia, 46100 Burjassot, Spain\\
$^8$ \quad Departamento de Física Matemática, Instituto de Física, Universidade de São Paulo, 05508-090, São
Paulo, SP, Brazil\\
$^9$ \quad Observatório Nacional, 20921-400, Rio de Janeiro, RJ, Brazil\\
$^{10}$\quad Instituto de Astrofísica de Andalucía - CSIC, 18080, Granada, Spain\\
$^{11}$\quad Donostia International Physics Center (DIPC), 20018 San Sebastián, Spain\\
$^{12}$\quad Centro de Estudios de Física del Cosmos de Aragón (CEFCA), 44001, Teruel, Spain\\
$^{13}$\quad Unidad Asociada CEFCA-IAA, CEFCA, Unidad Asociada al CSIC por el IAA y el IFCA, Plaza San Juan 1, 44001 Teruel, Spain\\
$^{14}$\quad Instituto de Astrofísica de Canarias, 38205, La Laguna, Tenerife, Spain \\
$^{15}$\quad National Astronomical Observatory of China,  Chinese Academy of Sciences, Beijing 100012, China\\
$^{16}$\quad Departamento de Astronomia, Instituto de Astronomia, Geofísica e Ciências Atmosféricas, Universidade de São Paulo, 05508-090, São Paulo, SP, Brazil\\
$^{17}$\quad Instruments4, La Canada Flintridge CA, 91011, USA\\
$^{18}$\quad Steward Observatory, University of Arizona, Tucson AZ, 85721, USA
}

\corres{Correspondence: valerio.marra@me.com}

\secondnote{These authors contributed equally to this work.}

\abstract{We present a supervised machine learning classification of sources from the Javalambre Physics of the Accelerating Universe Astrophysical Survey (J-PAS) Pathfinder datasets: miniJPAS and J-NEP. Leveraging crossmatches with spectroscopic and photometric catalogs, we construct a robust labeled dataset comprising 14\,594 sources classified into extended (galaxies) and point-like (stars and quasars) objects. We assess dataset representativeness using UMAP analysis, confirming broad and consistent coverage of feature space.
An XGBoost classifier, with hyperparameters tuned using automated optimization, is trained using purely photometric data (60-band J-PAS magnitudes) and combined photometric and morphological features, with performance thoroughly evaluated via ROC and purity–completeness metrics.
Incorporating morphology significantly improves classification, outperforming the baseline classifications available in the catalogs. Permutation importance analysis reveals morphological parameters, particularly concentration, normalized peak surface brightness, and PSF, alongside photometric features around 4000 and 6900 Å, as crucial for accurate classifications. We release a value-added catalog with our models for star-galaxy classification, enhancing the utility of miniJPAS and J-NEP for subsequent cosmological and astrophysical analyses.}


\begin{document}


\section{Introduction}

The Javalambre Physics of the Accelerating Universe Astrophysical Survey\footnote{\url{https://www.j-pas.org}} (J-PAS) is a ground-based imaging survey that will map thousands of square degrees using a unique set of 56 filters, providing a low-resolution spectrum for every pixel of the sky ($R \sim 60$),\footnote{The wavelength resolution $R_{\lambda}$ is defined as $R_{\lambda}=\lambda/\Delta \lambda$, so that $R \sim 60$ is the approximate value in the intermediate wavelength range in the J-PAS filter system.} without the target selection biases or fiber collision issues that typically affect spectroscopic surveys. Designed and optimized to achieve highly accurate photometric redshifts, its scientific objectives span from cosmology to galaxy evolution~\citep{Bonoli:2020ciz}.

Here we focus on data from the Pathfinder camera, which was used to test the telescope performance and execute the first scientific operations. The camera features a 9k~$\times$9k CCD, with a 0.3deg$^2$ field of view and 0.225~arcsec pixel size.
This effort led to the miniJPAS \citep{Bonoli:2020ciz} and J-NEP \citep{Hernan-Caballero:2023kaa} surveys, which covered $1\,\mathrm{deg}^2$ of the AEGIS field (four pointings around $(\mathrm{RA},\mathrm{Dec})=(215^\circ,53^\circ)$) and $0.29\,\mathrm{deg}^2$ of the North Ecliptic Pole (one pointing centred at $(\mathrm{RA},\mathrm{Dec})=(260.6^\circ,65.8^\circ)$), respectively.
Both surveys use the 56 J-PAS filters together with the $u$, $g$, $r$, and $i$ broad bands\footnote{The J-PAS survey itself employs only the $i$ broad band in addition to the 56 narrow bands.}, for a total of 60 bands. However, J-NEP adopted longer exposure times, reaching between 0.5--1.0 magnitudes deeper than miniJPAS, which itself reaches the target depth of J-PAS--namely, $m_{\rm AB} \approx 24$ in the broad-band filters (5$\sigma$ in a 3'' aperture).\footnote{Here $m_{\rm AB}$ denotes AB magnitudes, defined such that a source with constant flux density of $3631\,\mathrm{Jy}$ has $m_{\rm AB}=0$.}

While miniJPAS targeted the well-studied Extended Groth Strip field, an area observed by many experiments, J-NEP fully covers the time-domain field of the James Webb Space Telescope’s North Ecliptic Pole~\citep{2018PASP..130l4001J}, a region chosen for its low galactic extinction, absence of bright stars, and continuous visibility from Webb’s northern hemisphere.

Our main goal here is to provide a machine learning-based classification of J-NEP sources as an alternative to the Bayesian stellar and galactic loci classifier (SGLC \citep{2019A&A...622A.177L}, in \mytt{jnep.StarGalClass}\footnote{Throughout this work, for reproducibility, we explicitly reference the table names available at \href{https://www.j-pas.org/datareleases}{j-pas.org/datareleases}.}). This aims both to validate the quality of the SGLC, possibly outperfoming it, and to offer an independent classification that can be used to assess the robustness of subsequent scientific analyses. In particular, our approach incorporates the full photometric information, which is crucial for faint galaxies whose morphology closely resembles that of stars.

The SGLC is based on morphological information, specifically on the bimodal distribution observed in the concentration–magnitude diagram, which distinguishes compact point-like objects from extended sources. \citet{2019A&A...622A.177L} modeled these two populations to compute the probability that a given source is compact or extended. This model, combined with appropriate priors, was then used to estimate the Bayesian probability that a source is a star or a galaxy \citep{Bonoli:2020ciz}.

In \citet{Baqui:2020sfd}, we obtained the machine learning classification of miniJPAS sources, available in \mytt{minijpas.StarGalClass}. Here, taking advantage of new spectroscopic data and an improved methodology, we revise the classification of miniJPAS sources and provide a machine learning classification of J-NEP sources.
This paper is organized as follows. Section~\ref{sec:data} describes the training data, Section~\ref{sec:features} presents the feature configuration, Section~\ref{sec:ml} details the classification methodology and examines the representativeness of the labeled set, and Section~\ref{sec:res} reports the results. Finally, conclusions are drawn in Section~\ref{sec:conclu}.

\section{Data}
\label{sec:data}

In this work, we primarily used the \mytt{jnep.MagABDualObj} and \mytt{minijpas.MagABDualObj} tables from the data releases J-NEP-PDR202107 and MINIJ-PAS-PDR201912, respectively, available at \href{https://www.j-pas.org/datareleases}{j-pas.org/datareleases}. These catalogs provide photometric and morphological measurements for all sources detected in the $r$ band, with forced photometry in the remaining filters. The $r$ band was adopted as the detection image owing to its depth, image quality, and completeness, as detailed in \citet[][]{Bonoli:2020ciz}.
All queries employed in this work are listed in Appendix~\ref{ap:queries}.

To ensure data quality, we required objects to satisfy \mytt{mask_flags = 0}, which excludes detections affected by image boundaries, bright-star halos, or artifacts.  
For the \mytt{flags} field, which encodes issues encountered during photometric extraction, we retained objects with values \mytt{0} (clean, no problems detected), \mytt{1} (neighbor, the object is close to bright neighbors or contains bad pixels), \mytt{2} (blended, originally detected together with another source but successfully deblended), \mytt{4} (saturated, containing one or more pixels at or near saturation), and \mytt{6} (blended + saturated, a combination of the previous two cases). These situations can still yield reliable photometry and remain relevant for galaxy and cluster science, whereas more severe conditions (e.g., truncation at image edges, incomplete aperture measurements, or failed deblending) were excluded.
This yields a total of 25,198 objects for miniJPAS and 7,290 objects for J-NEP within the range $15 < r < 23.5$.

Since training supervised machine learning models requires ground-truth labels, we cross-match miniJPAS and J-NEP data with surveys that provide reliable photometric or spectroscopic classifications into extended (galaxies) and point-like (stars and quasars) sources. This is a crucial step, as the performance of a supervised ML model depends critically on the quality and representativeness of the adopted training set.
The surveys used for crossmatching in this work are:
\begin{itemize}

\item \textbf{SDSS DR12} (Section~\ref{sec:sdss}): the final Data Release (DR12) of the third-generation Sloan Digital Sky Survey \citep[SDSS-III,][]{2015ApJS..219...12A}, which collected imaging and spectroscopy from 2008 to 2014, providing wide-area multiband photometry and spectra of millions of sources.  

\item \textbf{Gaia EDR3} (Section~\ref{sec:gaia}): the Early Data Release 3 of the Gaia mission \citep{2021A&A...649A...1G}, delivering high-precision astrometry, broad-band photometry, and radial velocities for over 1.5 billion stars in the Milky Way.

\item \textbf{DEEP3 DR4} (Section~\ref{sec:deep3}): spectroscopic redshift survey conducted with DEIMOS on the Keck II telescope in the Extended Groth Strip (EGS) \citep{2011ApJS..193...14C}. It provides secure galaxy redshifts and spectra, forming part of the AEGIS survey program. 

\item \textbf{Binospec} (Section~\ref{sec:binospec}): spectroscopic data from Binospec~\citep{2019PASP..131g5004F}, a high-throughput optical spectrograph on the 6.5-m MMT, covering 370–1000~nm with wide field of view, used to obtain redshifts for faint galaxies.

\item \textbf{DESI DR1} (Section~\ref{sec:desi}): the first data release of the Dark Energy Spectroscopic Instrument \citep{DESI:2025fxa}, based on the first 13 months of the main survey. It provides high-confidence redshifts for 18.7 million objects (13.1M galaxies, 1.6M quasars, 4.0M stars) over more than 9000~deg$^2$, making it the largest extragalactic spectroscopic sample to date and a key resource for cosmology and large-scale structure studies.

\item \textbf{HSC-SSP PDR2} (Section~\ref{sec:subaru}): the second public data release of the Hyper Suprime-Cam Subaru Strategic Program \citep[HSC-SSP,][]{2019PASJ...71..114A}, providing deep, wide-field imaging in five broad bands ($grizy$), with exquisite seeing from the Subaru telescope.  

\end{itemize}

\subsection{Crossmatch with SDSS DR12}
\label{sec:sdss}

We used the table \mytt{minijpas.xmatch_sdss_dr12}, which contains the crossmatch with SDSS DR12 and includes both the photometric classification \mytt{class}, expected to be reliable for $15 < r < 20$ \citep{Baqui:2020sfd}\footnote{We verified this by comparing DESI, DEEP3, SDSS (spectroscopic), and Gaia sources (taken as ground truth) against the SDSS photometric classification.}, and the spectroscopic classification \mytt{spCl}.
The photometric classification yields 802 galaxies and 1228 stars within $15 < r < 20$, for a total of 2030 labeled sources. The spectroscopic classification provides 355 galaxies and 220 point-like sources (94 stars and 124 quasars) for $r <23.5$, for a total of 573 labeled sources.

We also used the analogous table \mytt{jnep.xmatch_sdss_dr12}. The photometric classification yields 156 galaxies and 724 stars within $15 < r < 20$, for a total of 880 labeled sources. No spectroscopic labels were available, since J-NEP lies outside the SDSS footprint.

\subsection{Crossmatch with Gaia EDR3}
\label{sec:gaia}

We used the tables \mytt{minijpas.xmatch_gaia_edr3} and \mytt{jnep.xmatch_gaia_edr3}, which contain the crossmatches with Gaia EDR3\footnote{As we are interested in parallax, Gaia EDR3 is essentially equivalent to the newer Gaia DR3.}. Stars were identified as objects with a parallax measured at a significance greater than 3$\sigma$, i.e., \mytt{parallax_over_error} > 3. This selection yields 989 stars for miniJPAS and 570 stars for J-NEP.

\subsection{Crossmatch between miniJPAS and DEEP3 DR4}
\label{sec:deep3}

\begin{figure}
\centering 
\includegraphics[trim={0 0 12cm 0}, clip, width= \columnwidth]{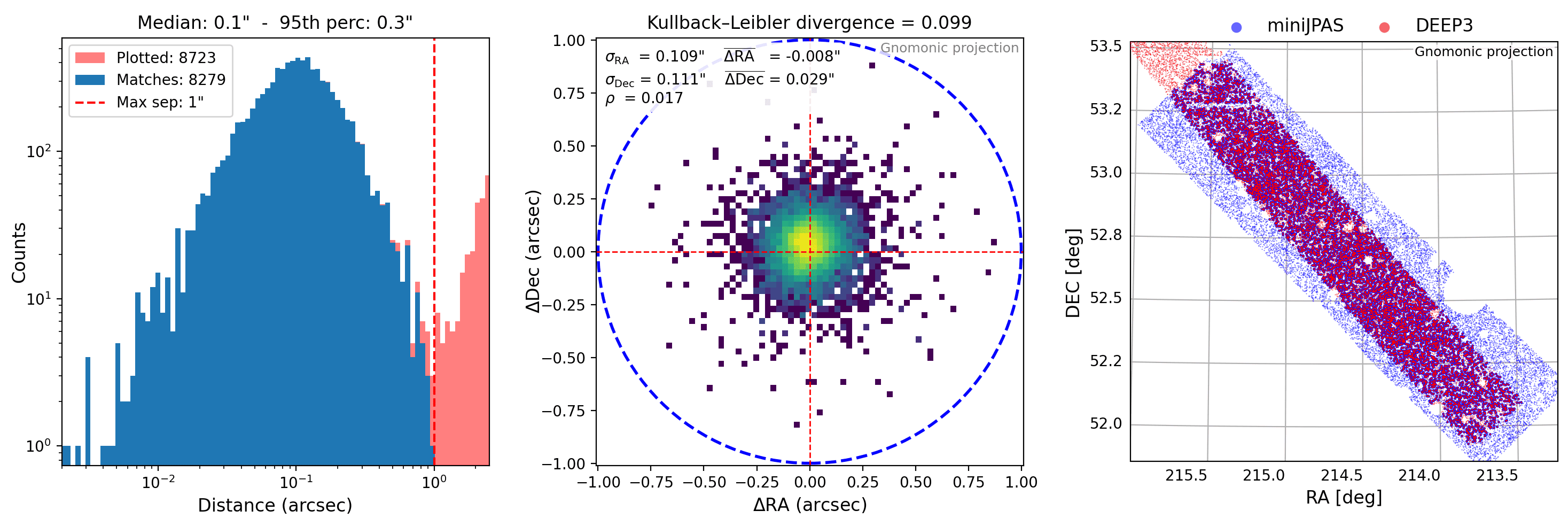}
\caption{Crossmatch between DEEP3 DR4 and miniJPAS sources with $r < 23.5$.}
\label{fig:xmatch_minijpas-deep3}
\end{figure}

We crossmatched miniJPAS with $r<23.5$ with the DEEP3 DR4 catalog \mytt{alldeep.egs.2012jun13}.\footnote{\url{https://sites.uci.edu/deep3/deep3zcat/}}
In the case of duplicates, we retained the entry with the highest \mytt{ZQUALITY}, and, if necessary, the one with the smallest redshift error \mytt{Z_ERR}. 
The crossmatch was performed using the Python module \mytt{onexmatch}\footnote{\url{https://github.com/valerio-marra/onexmatch}} \citep{marra_2025_17148385},
adopting a maximum separation of 1 arcsec and an ambiguity threshold of 0.5 arcsec, yielding 8279 matched objects as shown in Figure~\ref{fig:xmatch_minijpas-deep3}.

As shown by the diagnostic metrics, the positional offsets in Right Ascension and Declination are well described by a zero-mean isotropic 2D Gaussian whose standard deviation is equal to the average of the measured $\sigma_{\Delta \mathrm{RA}}$ and $\sigma_{\Delta \mathrm{DEC}}$; the resulting Kullback–Leibler divergence is  $\lesssim 0.1$. The adoption of an ambiguity threshold of 0.5 arcsec removed 10 ambiguous DEEP3 matches associated with duplicated miniJPAS sources, whose positional differences were below this threshold—about twice the pixel scale and comparable to the PSF size. See Appendix~\ref{ap:ambiguous} for more details.

After crossmatching, we retained only sources with \mytt{ZQUALITY>=3}, yielding 6583 galaxies, 433 active galactic nuclei (AGNs), and 5 stars. Unlike quasars (QSOs), many AGNs exhibit extended morphologies, which introduces ambiguity in the classification process. To mitigate this, we restricted point-like sources to objects with \mytt{morph_prob_star}>$0.9$ and redshift \mytt{Z}>$0.1$, and stars to those with \mytt{morph_prob_star}>$0.9$ and \mytt{Z}<$0.01$. The \mytt{morph_prob_star} values are provided by the \mytt{jnep.StarGalClass} table, which estimates stellar probability from morphological information \citep{2019A&A...622A.177L}. This selection resulted in 36 point-like sources (35 AGNs and 1 star).
This filtering step was necessary to construct a clean labeled dataset. Our study focuses on a binary star-galaxy classification, rather than a three-class star-AGN/QSO-galaxy scheme, which would require a significantly larger and more balanced AGN sample.

\subsection{Crossmatch between J-NEP and Binospec data}
\label{sec:binospec}

\begin{figure}
\centering 
\includegraphics[trim={0 0 12cm 0}, clip, width= \columnwidth]{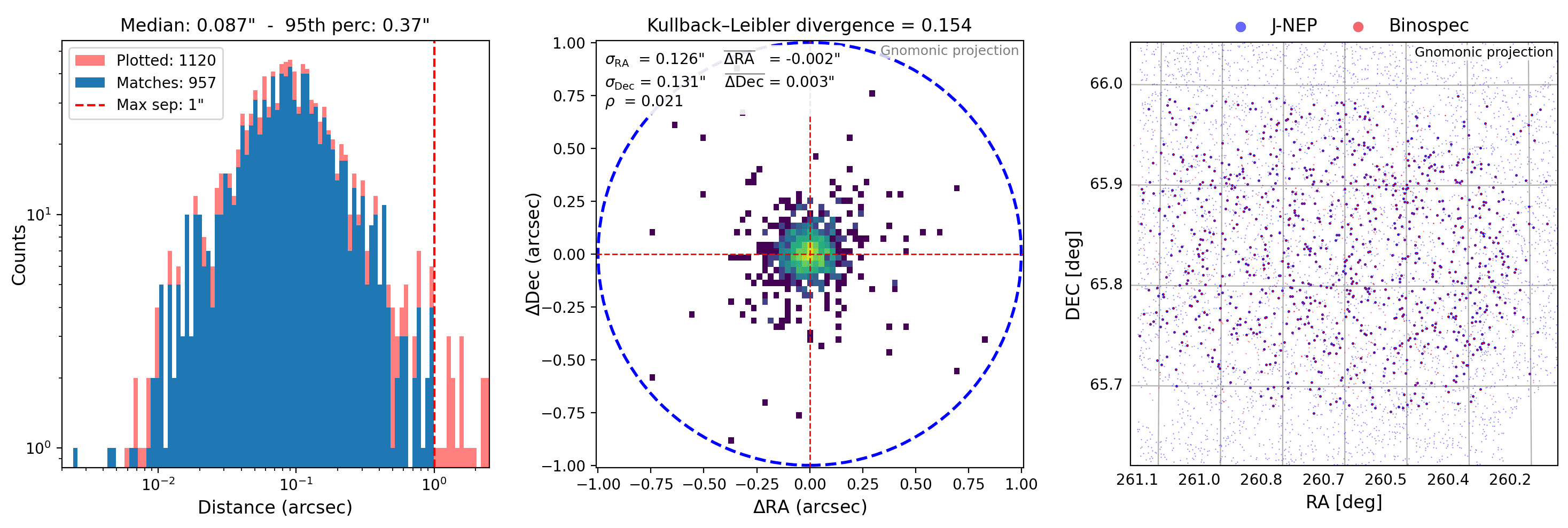}
\caption{Crossmatch between Binospec and J-NEP sources with $r < 23.5$.}
\label{fig:xmatch_jnep-binospec}
\end{figure}

We crossmatched J-NEP sources with $r<23.5$ with data that was obtained by Christopher Willmer using Binospec at the MMT observatory. Details about the dataset are provided in \citet{Hernan-Caballero:2023kaa}.
The crossmatch was performed using the \mytt{onexmatch} code, adopting a maximum separation of 1 arcsec and an ambiguity threshold of 0.5 arcsec, and yielded 957 matched sources, as shown in Figure~\ref{fig:xmatch_jnep-binospec}.  The adoption of an ambiguity threshold of 0.5 arcsec removed 67 ambiguous Binospec matches associated with duplicated miniJPAS sources, mostly with positional differences below 0.2 arcsec (Figure~\ref{fig:ambiguous}).

After crossmatching, we retained only the 726 objects with \mytt{ZQUALITY >= 3} and \mytt{Z_ERR < 0.01}.
Since the Binospec catalog does not provide spectroscopic classifications, we assigned class labels based on morphology and redshift: objects with \mytt{morph_prob_star} $> 0.9$ and $Z < 0.01$ were classified as stars, and those with \mytt{morph_prob_star} $< 0.1$ and $Z > 0.01$ as galaxies.  This yielded 53 stars and 593 galaxies, for a total of 646 objects.
There are 69 objects with \mytt{morph_prob_star} $> 0.1$ and $Z > 0.01$. These could correspond either to QSOs or to compact galaxies; we exclude this small set of sources from the catalog.

\subsection{Crossmatch between miniJPAS and DESI DR1}
\label{sec:desi}

\begin{figure}
\centering 
\includegraphics[trim={0 0 12cm 0}, clip, width= \columnwidth]{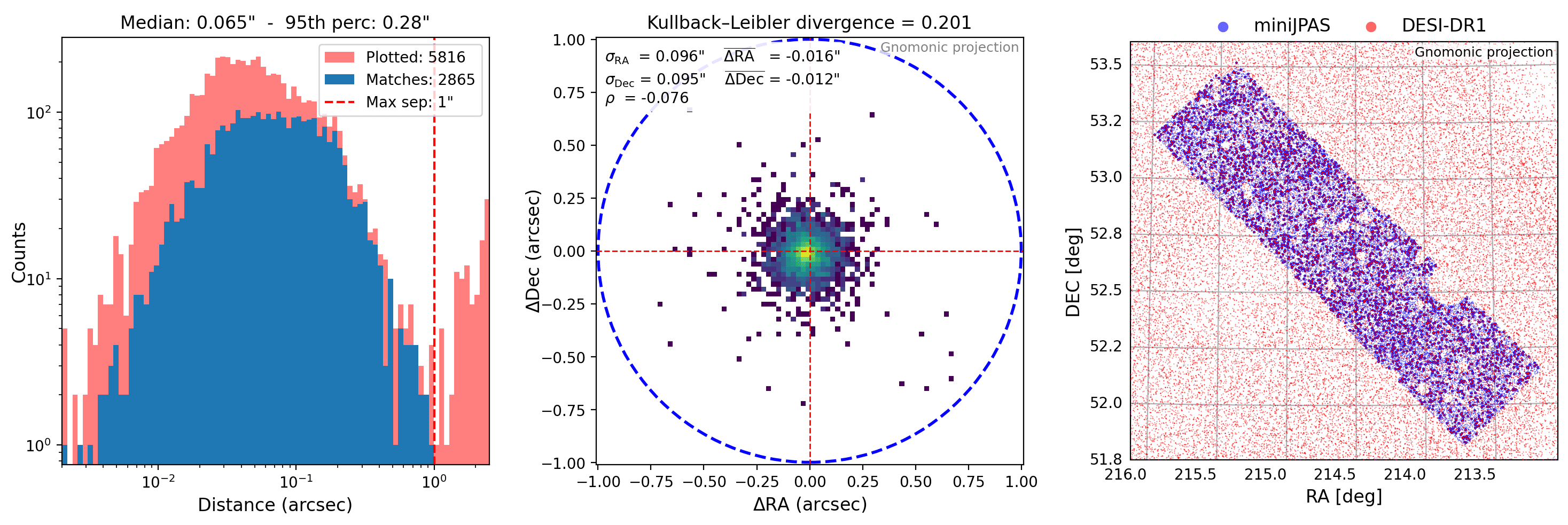}
\caption{Crossmatch between DESI DR1 and miniJPAS sources with $r < 23.5$.}
\label{fig:xmatch_minijpas-desi}
\end{figure}

We crossmatched miniJPAS sources with $r<23.5$ with the DESI DR1 catalog.\footnote{\mytt{zall-pix-iron.fits} at \url{https://data.desi.lbl.gov/public/dr1/spectro/redux/iron/zcatalog/v1}}
The crossmatch was performed using the \mytt{onexmatch} code, adopting a maximum separation of 1 arcsec and an ambiguity threshold of 0.5 arcsec, and yielded 2865 matches, as shown in Figure~\ref{fig:xmatch_minijpas-desi}. The adoption of an ambiguity threshold of 0.5 arcsec removed 1232 ambiguous DESI matches associated with duplicated miniJPAS sources, mostly with positional differences below 0.05 arcsec (Figure~\ref{fig:ambiguous}).

After crossmatching, as suggested in \citep{DESI:2025fxa}, only sources with \mytt{ZCAT_PRIMARY=True}, \mytt{ZWARN=0} and \mytt{DELTACHI2>25} were retained, for a total of 2640 objects.
Of these, 2307 are galaxies and 333 are point-like sources (124 QSOs and 209 stars).

\subsection{Crossmatch between miniJPAS and HSC-SSP PDR2}
\label{sec:subaru}

We used the photometric classification provided by \mytt{x_extendedness_value} (with \mytt{x=g, r, i, z, y}) from the catalog \mytt{minijpas.xmatch_subaru_pdr2}, which contains the crossmatch with the HSC-SSP PDR2 catalog. 
To ensure consistency, we retained only the 11\,960 objects for which all five \mytt{x_extendedness_value} measurements agree. 

\begin{table}
\centering
\setlength{\tabcolsep}{5pt}
\caption{Confusion matrices in three $r$-band magnitude bins, comparing DESI, DEEP3, SDSS (spectroscopic) and Gaia sources (used as ground truth) against HSC-SSP classifications.}\label{tab:subaru}
\begin{tabular}{c ccc ccc ccc}
\toprule
 & \multicolumn{2}{c}{18.5--20.5} & & \multicolumn{2}{c}{20.5--22.5} & & \multicolumn{2}{c}{22.5--23.5} \\
\cmidrule{2-3} \cmidrule{5-6} \cmidrule{8-9}
 & pred.\ stars & pred.\ gal. & & pred.\ stars & pred.\ gal. & & pred.\ stars & pred.\ gal. \\
\midrule
true stars & 79 & 10 & & 92 & 21 & & 6 & 4 \\
true gal. & 0 & 385 & & 5 & 1523 & & 9 & 2350 \\
\bottomrule
\end{tabular}
\end{table}

We tested the HSC-SSP classification in the range $18.5 < r < 23.5$ against DESI, DEEP3, SDSS (spectroscopic) and Gaia sources, used as ground truth; see Table~\ref{tab:subaru}. We find that for $r>22.5$ a non-negligible fraction of stars, though small in absolute numbers, are misclassified as galaxies.  
To minimize this effect, we restrict the HSC-SSP labels to the range $18.5 < r < 22.5$, where the bright limit avoids saturation.  
This selection yields 4406 galaxies and 946 stars, for a total of 5352 objects.

\subsection{Final labeled set}

\begin{table}
\centering
\setlength{\tabcolsep}{10pt}
\renewcommand{\arraystretch}{1.2}
\begin{tabular}{lcc}
\toprule
Crossmatch Source        & miniJPAS (galaxies/stars) & J-NEP (galaxies/stars) \\
\midrule
DESI DR1                 & 2307/333 (2307/333)        & --                     \\
Binospec                & --        & 593/53 (593/53)               \\
DEEP3 DR4                & 5748/16 (6583/36)        & --                     \\
SDSS DR12 (spectro)      & 189/124 (355/218)        & --                     \\
Gaia EDR3                & --/826 (--/989)        & --/565 (--/570)                \\
HSC-SSP PDR2             & 2323/765 (4406/946)        & --                     \\
SDSS DR12 (photo)        & 157/261 (802/1228)        & 105/229 (156/724)              \\
\midrule
Total                    & 10724/2325        & 698/847              \\
\bottomrule
\end{tabular}
\caption{Summary of external catalogs used to crossmatch and label sources in miniJPAS and J-NEP. Values in parentheses indicate the total number of matched sources before duplicate removal. When a source appears in multiple catalogs, labels are assigned based on a priority order corresponding to the table’s top-down sequence, with DESI given the highest priority. As a result, the final number of adopted labels is smaller than the per-catalog match count.}
\label{tab:xmatch_summary}
\end{table}

\begin{figure}
\centering 
\includegraphics[width= \columnwidth]{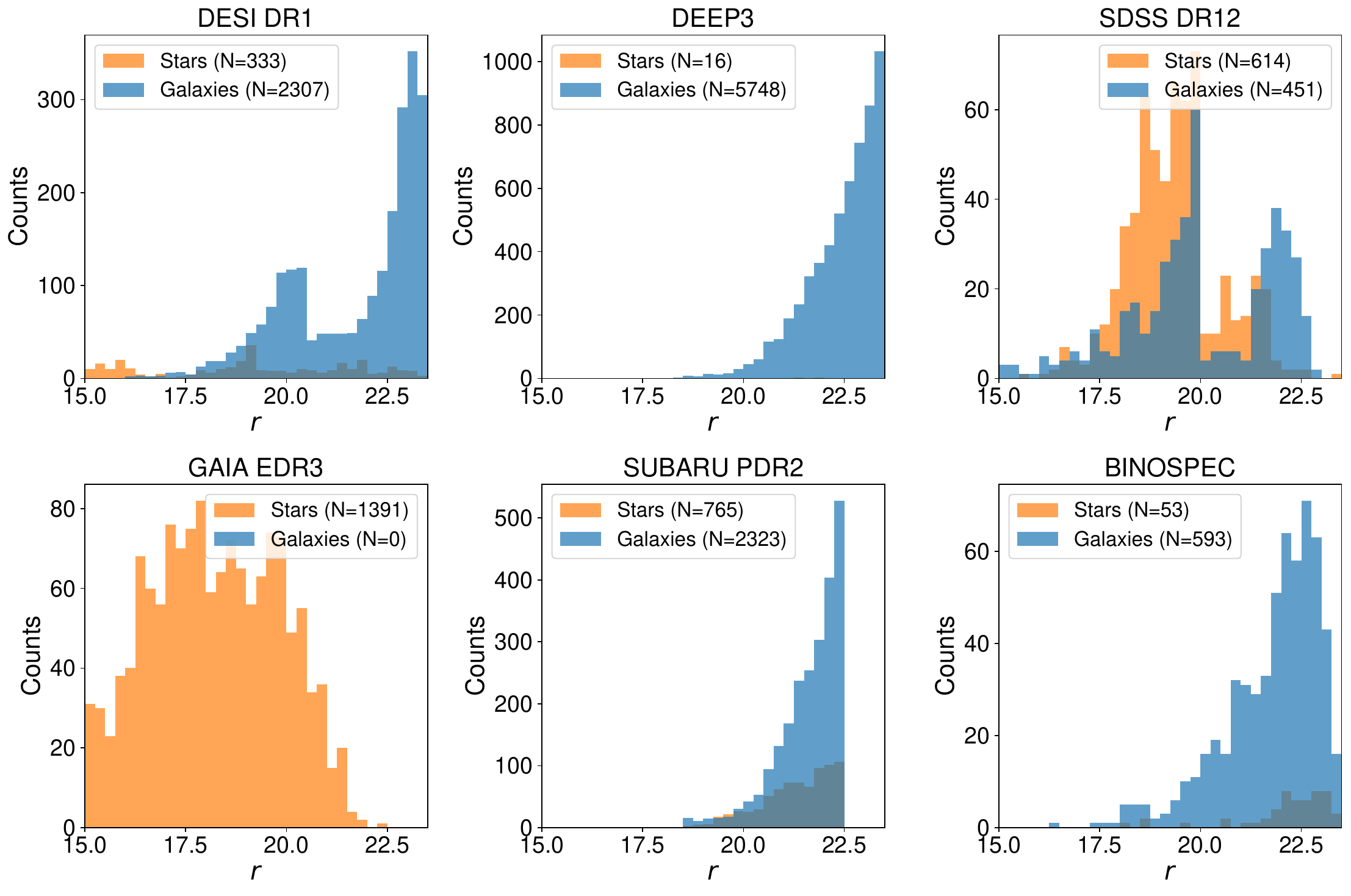}
\caption{$r$-band magnitude distributions of labeled stars and galaxies across the catalogs used for crossmatching. SDSS spectroscopic and photometric classifications are separated at $r=20$.}
\label{fig:surveys-mags}
\end{figure}

The crossmatched catalogs from the previous sections partially overlap. When dealing with duplicate J-PAS sources, we assigned class labels according to the following priority order, reflecting the reliability of each classification: DESI, Binospec, DEEP3, SDSS (spectroscopic classification), Gaia, HSC-SSP, and SDSS (photometric classification). Table~\ref{tab:xmatch_summary} summarizes the labeled dataset, comprising 14\,594 unique sources. The corresponding $r$-band magnitude distributions of the retained sources from each survey are shown in Figure~\ref{fig:surveys-mags}.

\begin{figure}
\centering 
\includegraphics[width= .8\columnwidth]{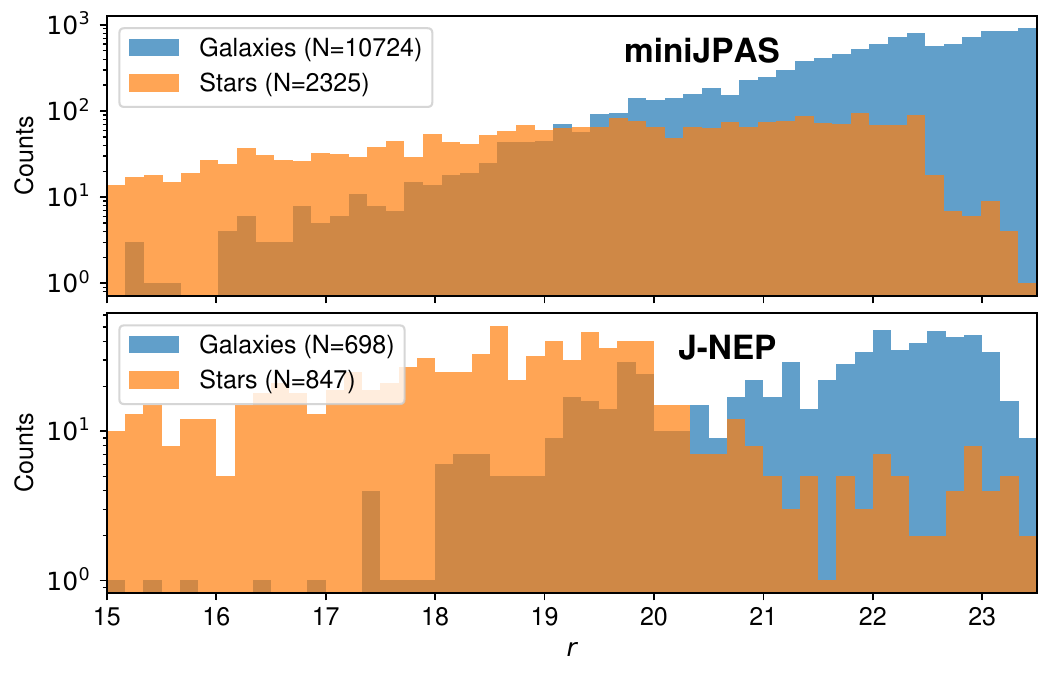}
\caption{$r$-band magnitude distributions of labeled stars and galaxies in miniJPAS and J-NEP.}
\label{fig:jnep-minijpas-mags}
\end{figure}

Figure~\ref{fig:jnep-minijpas-mags} shows the $r$-band magnitude distributions of stars and galaxies for the miniJPAS and J-NEP surveys. In both surveys, galaxies dominate at $r \gtrsim 21$, while stars are more prevalent at $r \lesssim 18.5$. This results in a class imbalance in the training set, which can bias both the performance of the classifier and the interpretation of its metrics. To avoid misleading conclusions from this imbalance, we evaluate performance using purity-completeness curves, which provide a balanced view of classification quality independent of the relative proportions of stars and galaxies.

\begin{figure}
\centering
\includegraphics[width=\columnwidth]{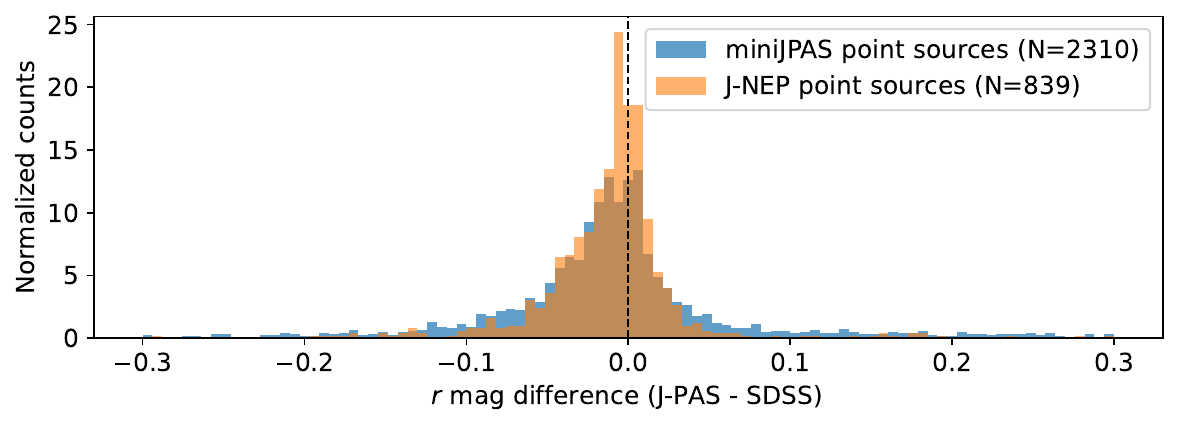}
\caption{Histogram of the differences between the SDSS $r$-band model magnitudes and the J-PAS $r$-band \mytt{MAG_APER_COR_3} magnitudes for the stars of the final labeled set. The latter corresponds to the total magnitude obtained from a 3 arcsec diameter aperture corrected for aperture effects.}
\label{fig:jnep-minijpas-sdss}
\end{figure}

Finally, Figure~\ref{fig:jnep-minijpas-sdss} shows the histogram of the magnitude differences between the SDSS and J-PAS $r$-band total flux magnitudes for point sources. This provides an additional crosscheck of the quality of the crossmatch. We restrict the analysis to point sources because J-PAS does not provide a total-flux model magnitude for extended objects, as \mytt{MAG_AUTO} is measured within an aperture of two Kron radii and therefore misses a non-negligible fraction of the galaxy flux in the outskirts.

As stated earlier, we perform a binary classification into extended and point-like sources. Point-like sources include both stars and QSOs, which are morphologically similar at the J-PAS resolution. However, the number of spectroscopically confirmed QSOs in our sample is insufficient to support a reliable three-class classification \citep{vonMarttens:2022mpv,2023MNRAS.518.3123C, 2024A&A...691A.221D, 2024MNRAS.527.4677Z, 2025A&A...700A.259A} that would exploit the low-resolution J-PAS spectra. As is customary, we refer to extended sources as galaxies and to point-like sources as stars throughout this paper.

\section{Feature configuration}
\label{sec:features}

\subsection{Observational features}
\label{sec:obs-feat}

To characterize the observing conditions and data quality, we included four observational features:
\begin{enumerate}
\item Point Spread Function (PSF) of the tile,
\item Galactic extinction $E(B-V)$,
\item $E(B-V)$ error,
\item Object \mytt{flags} indicating observational issues.
\end{enumerate}
The PSF is characterized as the mean full width at half maximum (FWHM) of point-like sources in each tile, estimated with \mytt{PSFEx}~\citep{2011ASPC..442..435B}.
The color excess $E(B-V)$ is the integrated reddening due to Milky Way dust along the line of sight, provided at the position of each source by the Bayestar17 three-dimensional dust map \citep{2018MNRAS.478..651G,2019A&A...631A.119L}.

\subsection{Photometric features}

We used a total of 120 photometric features derived from the 60 J-PAS bands. Specifically, we considered \mytt{MAG_AUTO} magnitudes and their associated uncertainties in each of the 60 bands from \mytt{MagABDualObj}. Bands without detection were assigned a magnitude of 99, while the magnitude errors encapsulate valuable information about the strength and reliability of each detection.

Although \mytt{MagABDualPointSources} provides PSF-corrected photometry (\mytt{MAG_APER_COR_3_0}) optimized for point-like objects--shown to outperform \mytt{MAG_AUTO} for spectral energy distribution (SED) fitting and photometric redshift estimation--our classification tests indicate a different trend. For the photometry-only model, \mytt{MAG_AUTO} yields  better results, with ROC AUC = 0.992 and star purity-completeness AUC = 0.984, compared to 0.984 and 0.965 for \mytt{MAG_APER_COR_3_0}.\footnote{ROC stands for receiver operating characteristic, and AUC denotes the area under the curve.} For galaxies, the improvement is smaller but still in favor of \mytt{MAG_AUTO} (AUC = 0.997 vs.\ 0.994). When morphology is included, the two definitions perform nearly identically--0.1\% better performance for \mytt{MAG_APER_COR_3_0}.
Given the superior performance of \mytt{MAG_AUTO} in the photometry-only case and its comparable performance when morphology is included, we adopt \mytt{MAG_AUTO} as our default choice in this work.

\subsection{Morphological features}

\begin{figure}
\centering 
\includegraphics[trim={0 0 0 0}, clip, width= \columnwidth]{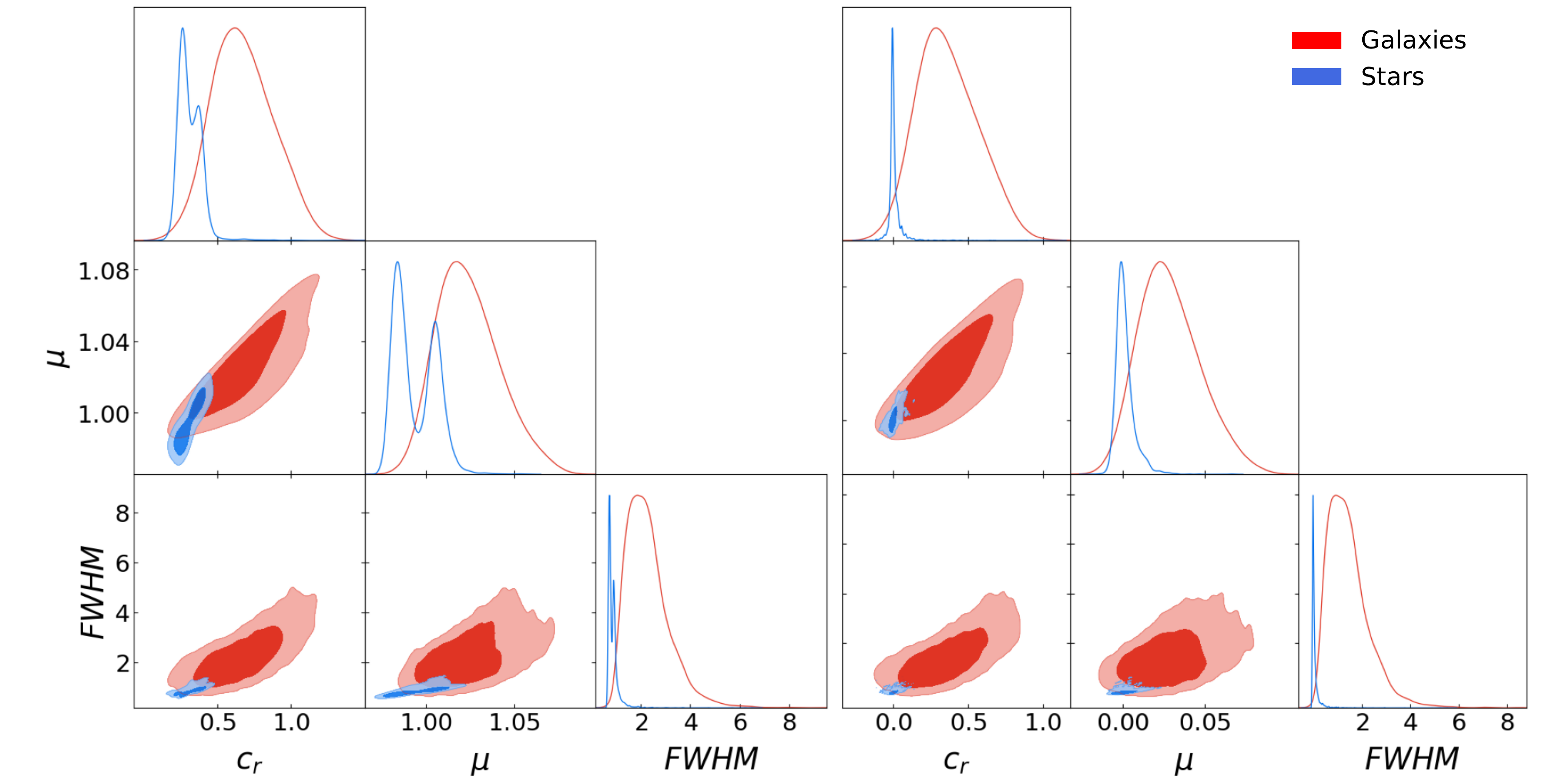}
\caption{Distribution of morphological features before (left) and after (right) subtracting the tile-by-tile median stellar values.
For each panel, the dark and light shades show the 68\% and 95\% contours of the distributions, respectively.
The original distributions (left) are affected by PSF-induced multimodality across tiles.}
\label{fig:triplot_morpho}
\end{figure}

We adopted four morphological parameters derived from columns relative to the detection $i$ band, available in the \mytt{MagABDualObj} catalog:
\begin{enumerate}
\item \textit{Ellipticity} $A/B = \mytt{A_WORLD}/\mytt{B_WORLD}$, where \mytt{A_WORLD} and \mytt{B_WORLD} are the root mean square values of the light distribution along the major and minor axes of the source, respectively.
\item \textit{Concentration} $c_r = \mytt{MAG_APER_1_5} - \mytt{MAG_APER_3_0}$ (in the $r$-band), calculated from magnitudes measured within circular apertures of 1.5 and 3.0 arcsec.
\item \textit{Full Width at Half Maximum} $\mytt{FWHM_WORLD}$, defined assuming a Gaussian intensity distribution.
\item \textit{Normalized peak surface brightness} $\mu=\mytt{MU_MAX}/\mytt{MAG_APER_3_0}$ (in the $r$-band), where \mytt{MU_MAX} represents the peak surface brightness above the local background.
\end{enumerate}

The spatial variation of the PSF across the J-PAS tiles introduces multimodality into the distributions of concentration, FWHM, and peak surface brightness. To correct for this effect and achieve uniform morphological features, we normalize each parameter by subtracting the corresponding median stellar value computed within the same tile:
\begin{align}
X_i \;\;\rightarrow\;\; X_i - \langle X \rangle_{\mathrm{stars,\,tile}} \, ,
\end{align}
where $X$ denotes concentration, FWHM, or peak surface brightness.
This correction effectively removes the multimodal behavior, as illustrated in Figure~\ref{fig:triplot_morpho}, which compares the distributions before and after correction. In addition, the separation between stellar and galactic populations becomes more pronounced following this normalization.

We do not apply this correction to the ellipticity, as it does not display multimodal features and can take spurious values for stars. Due to the finite pixel scale, a narrow PSF may lead SExtractor to assign $A/B \neq 1$ to unresolved sources, introducing artificial shape distortions caused by limitations in the PSF fitting procedure.
See \citet[][App.~C]{Bonoli:2020ciz} for the SExtractor configuration adopted.

We use the normalized morphological parameters for the representativeness analysis presented in the next section. However, for the classification task, we retain the uncorrected catalog values. Our tests show that including the PSF as an additional feature is sufficient to account for its spatial variations and prevent confusion, while the normalization procedure introduces additional complexity to the pipeline without yielding a significant improvement in classification performance.

\subsection{Feature sets}

We perform three separate classifications. The first uses the full set of 128 features, which includes 60 magnitudes with their associated errors, four morphological parameters, the PSF information, Galactic extinction with its error, and detection flags. The second is restricted to 123 features, consisting of the 60 magnitudes with their errors, Galactic extinction with its error, and detection flags. We refer to these two configurations as `morphology+photometry` and `photometry only`, respectively. The rationale for testing the latter is that morphological parameters may be biased in tiles produced by combining exposures taken under different observing conditions; in such cases, the tile-averaged PSF may not adequately represent individual sources, making a photometry-only classification potentially more robust.

In addition, we evaluate a third model based solely on morphological information, comprising the four morphological parameters, PSF, and detection flags. This setup serves as a benchmark to compare with the other models and, in particular, with the SGLC classifier, thereby quantifying the performance gain provided by machine learning.

\section{Supervised classification}
\label{sec:ml}

\subsection{Representativeness}
\label{repre}

\begin{figure}
\centering 
\includegraphics[trim={0 0 0 0}, clip, width= \columnwidth]{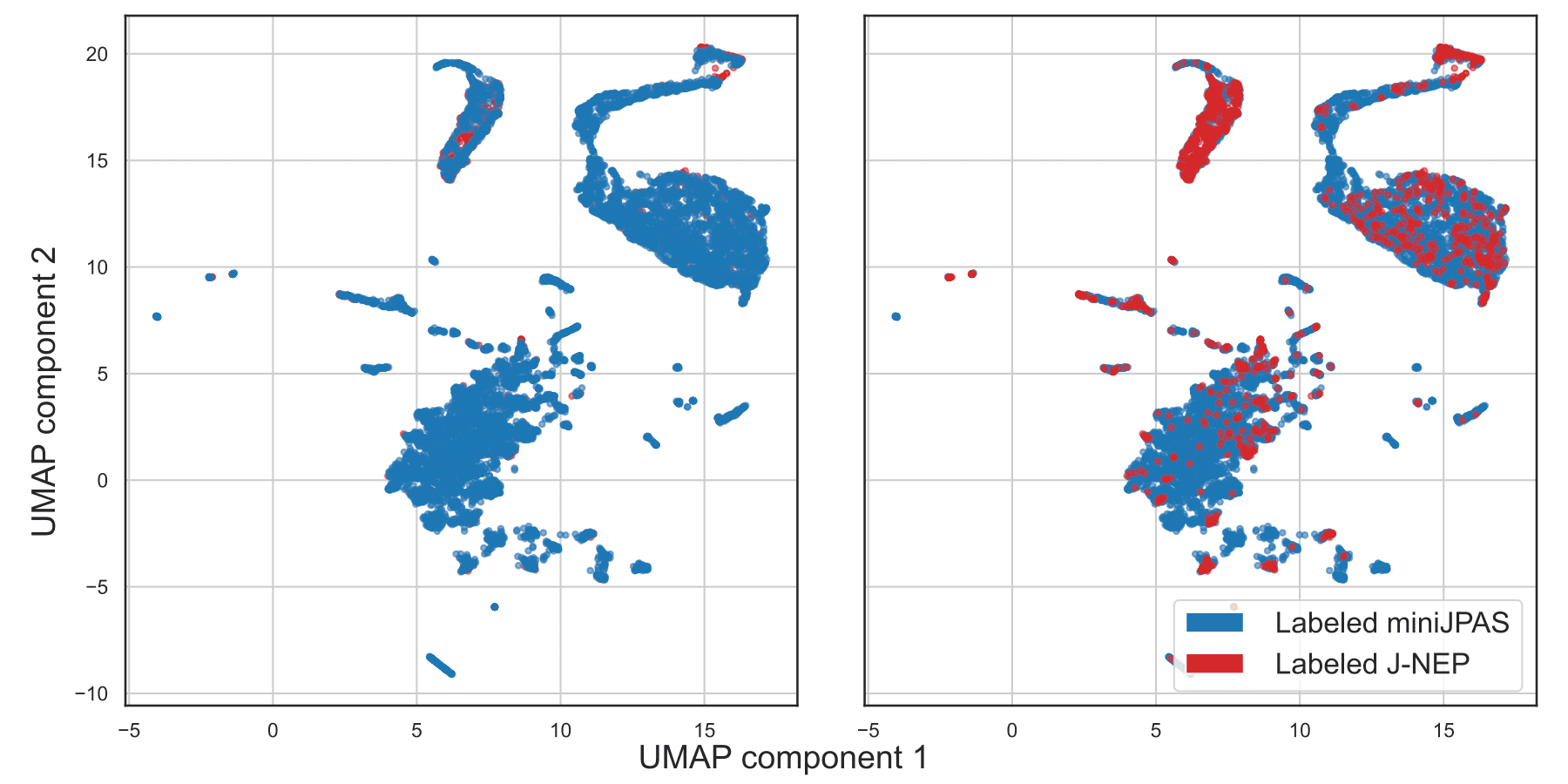}
\caption{UMAP projection of labeled miniJPAS and J-NEP samples, showing consistent coverage across feature space. Left (right) panel shows miniJPAS (J-NEP) sources plotted on top to enhance visibility.}
\label{fig:umap_labeled}
\end{figure}

The miniJPAS and J-NEP surveys were conducted using the same instrument and filter system, though under slightly different observing conditions. To evaluate the reliability and generalization capacity of the labeled datasets assembled in Section~\ref{sec:data}, we assess their representativeness along two complementary directions:
\begin{enumerate}
\item the internal consistency between the labeled miniJPAS and labeled J-NEP subsets;
\item the external coverage of the full miniJPAS and J-NEP datasets by the combined labeled set.
\end{enumerate}

For this purpose, we adopt the Uniform Manifold Approximation and Projection (UMAP) algorithm \citep{2018arXiv180203426M}, a non-linear dimensionality reduction technique designed to preserve both local and global data structure in low-dimensional projections. Importantly, our use of UMAP is diagnostic: we do not attempt classification in the embedded space, but rather evaluate whether the labeled samples span the same feature manifold as the full catalogs.

\begin{figure}
\centering 
\includegraphics[trim={0 0 0 0}, clip, width= \columnwidth]{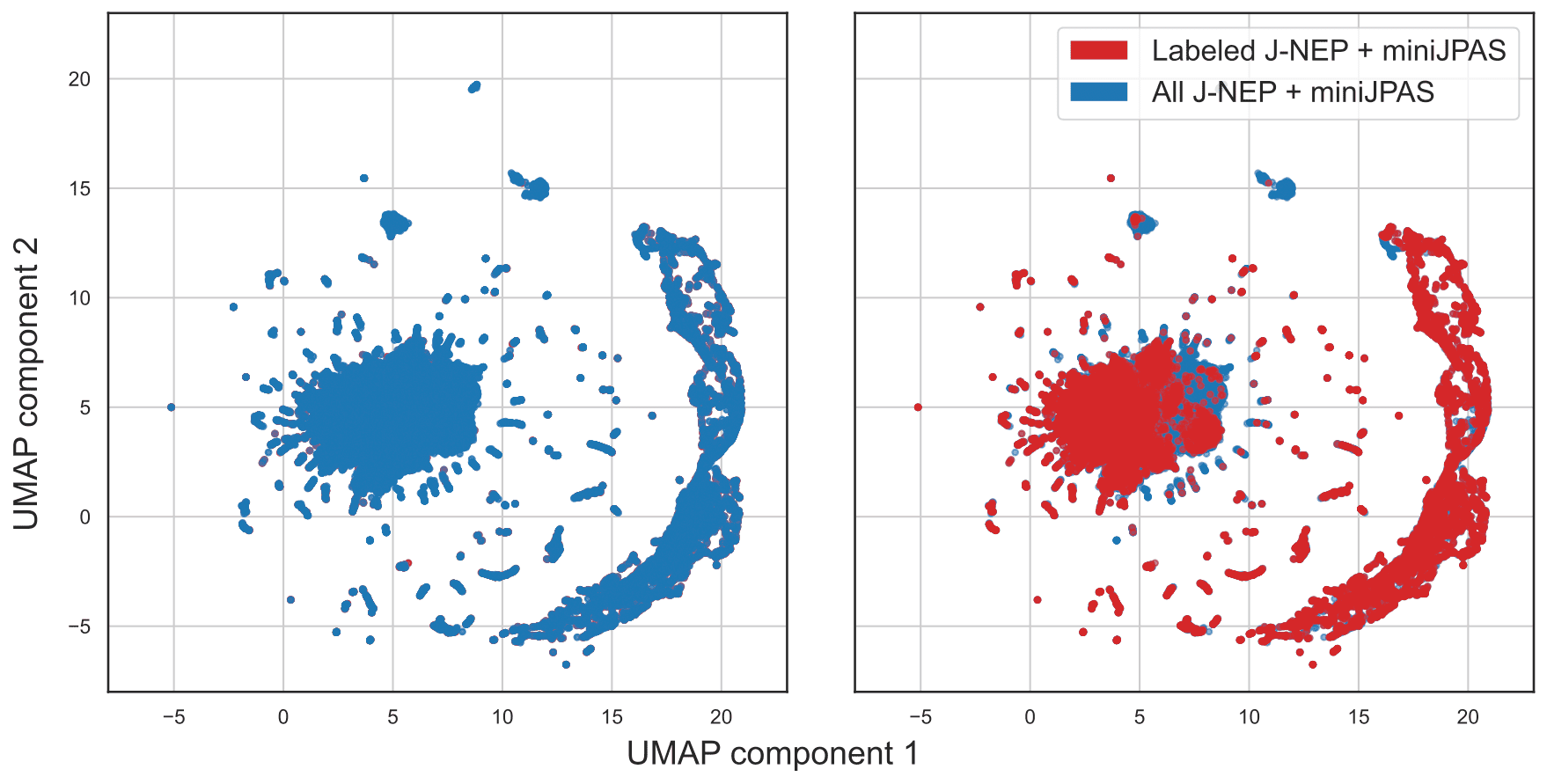}
\caption{UMAP projection of the combined labeled set and full miniJPAS and J-NEP catalogs, showing representative coverage. Left (right) panel shows miniJPAS (J-NEP) sources plotted on top to enhance visibility.}
\label{fig:umap_vac}
\end{figure}

We apply UMAP to the 124 photometric and morphological features described in Section~\ref{sec:features}, excluding the observational features listed in Section~\ref{sec:obs-feat}, as these are non–object-specific or discrete and could introduce spurious clustering.
Figure~\ref{fig:umap_labeled} shows the two-dimensional UMAP projection of the labeled miniJPAS and J-NEP sources. The J-NEP subset is largely embedded within the distribution of miniJPAS, indicating that the two labeled samples are consistent and can be reliably combined for model training.
Figure~\ref{fig:umap_vac} compares the combined labeled dataset against the full miniJPAS and J-NEP populations.\footnote{Note that the UMAP projection differs from that in Figure~\ref{fig:umap_labeled}, so the two plots are not directly comparable in terms of their coordinate positions.}
The labeled set broadly spans the full feature space, confirming that it is sufficiently representative to serve as a training base for supervised classification of the entire catalog.
It is important to note, however, that while the labeled set spans the overall feature space, this does not guarantee that labeled stars (or galaxies) fully cover the star (or galaxy) subspaces of the full catalogs.

\subsection{TPOT pipeline for XGBoost optimization}
\label{ssec:tpotxgb}

Based on the results of \citet{vonMarttens:2022mpv}, who compared several machine learning algorithms, we adopt the eXtreme Gradient Boosting algorithm \citep[XGBoost,][]{Chen:2016btl} for classification.
Given the relatively modest size of our labeled dataset (fewer than 15\,000 objects), XGBoost is a suitable choice. While neural networks may outperform tree-based models for substantially larger datasets, they are unlikely to offer significant advantages in our case.
XGBoost is an ensemble learning method based on decision trees and gradient boosting. It iteratively adds new trees to correct the errors of the existing ensemble while minimizing a regularized objective function, which balances prediction accuracy and model complexity. This process results in a robust and efficient classifier that is well-suited for structured tabular data. For technical details, see \citet{Chen:2016btl}.

The final hyperparameter configuration was selected using the Tree-based Pipeline Optimization Tool \citep[TPOT,][]{le2020scaling},\footnote{\url{https://epistasislab.github.io/tpot/latest/}} an automated machine learning framework that optimizes machine learning pipelines through genetic programming. The number of pipelines evaluated by TPOT depends on three key hyperparameters: \textit{generations}, \textit{population size}, and \textit{offspring size}. These respectively control the number of optimization iterations, the number of top-performing pipelines retained in each generation, and the number of new pipelines generated per generation.

The total number of pipelines analyzed is given by 
\textit{Population Size} + \textit{Generations} $\times$ \textit{Offspring Size}. 
In this work, we set all three hyperparameters to 200, resulting in the evaluation of 40,200 distinct pipelines. 
The final hyperparameter configuration is listed in Appendix~\ref{ap:hp}. 
Model selection was performed exclusively within the training set, comprising 80\% of the combined miniJPAS and J-NEP labeled samples.  
TPOT used 5-fold cross-validation within this training set to optimize pipeline architectures and hyperparameters.  
After optimization, the best-performing pipeline was retrained on the same 80\% split, while the remaining 20\% was held out as an independent test set. 
This test set was used only once for final evaluation, ensuring that the reported performance provides an unbiased estimate of the model’s generalization ability.

\subsection{Performance metrics}
\label{ssec:metrics}

The XGBoost classifier outputs a probabilistic score between 0 (star) and 1 (galaxy), requiring the choice of a classification threshold $p_{\rm cut}$ to separate the two classes. Once $p_{\rm cut}$ is specified, the model’s performance can be evaluated using a $2\times 2$ confusion matrix. From its entries, we derive standard classification metrics such as the receiver operating characteristic (ROC) curve and the purity–completeness curve, which provide graphical assessments of classifier performance.

The ROC curve is a parametric plot of the true positive rate (TPR) versus the false positive rate (FPR) as a function of $p_{\rm cut}$:
\begin{align}
\text{TPR}(p_{\rm cut}) = \frac{\rm TP}{\rm TP + FN}\,, \quad
\text{FPR}(p_{\rm cut}) = \frac{\rm FP}{\rm FP + TN} \,,
\end{align}
where TP, FN, FP, and TN are the entries of the confusion matrix. TPR corresponds to recall, commonly referred to as completeness in astronomy. A commonly used scalar summary of the ROC curve is the area under the curve (AUC), which ranges from 0.5 (random classifier) to 1 (perfect classifier).

To account for class imbalance in the training set, we also consider the purity-completeness curve. It is a parametric plot of the purity (or precision) versus the completeness (or recall) as a function of $p_{\rm cut}$:
\begin{align}
\text{Purity}(p_{\rm cut}) = \frac{\rm TP}{\rm TP + FP}\,, \quad
\text{Completeness}(p_{\rm cut}) = \frac{\rm TP}{\rm TP + FN} \,.
\end{align}
A compact summary of this curve is given by the average precision (AP), defined as the area under the purity–completeness curve. AP values range from 0 to 1, with higher values indicating better classification performance.

\section{Results}
\label{sec:res}

\subsection{Metrics}

\begin{figure}
\centering 
\includegraphics[trim={0 0 0 0}, clip, width= \columnwidth]{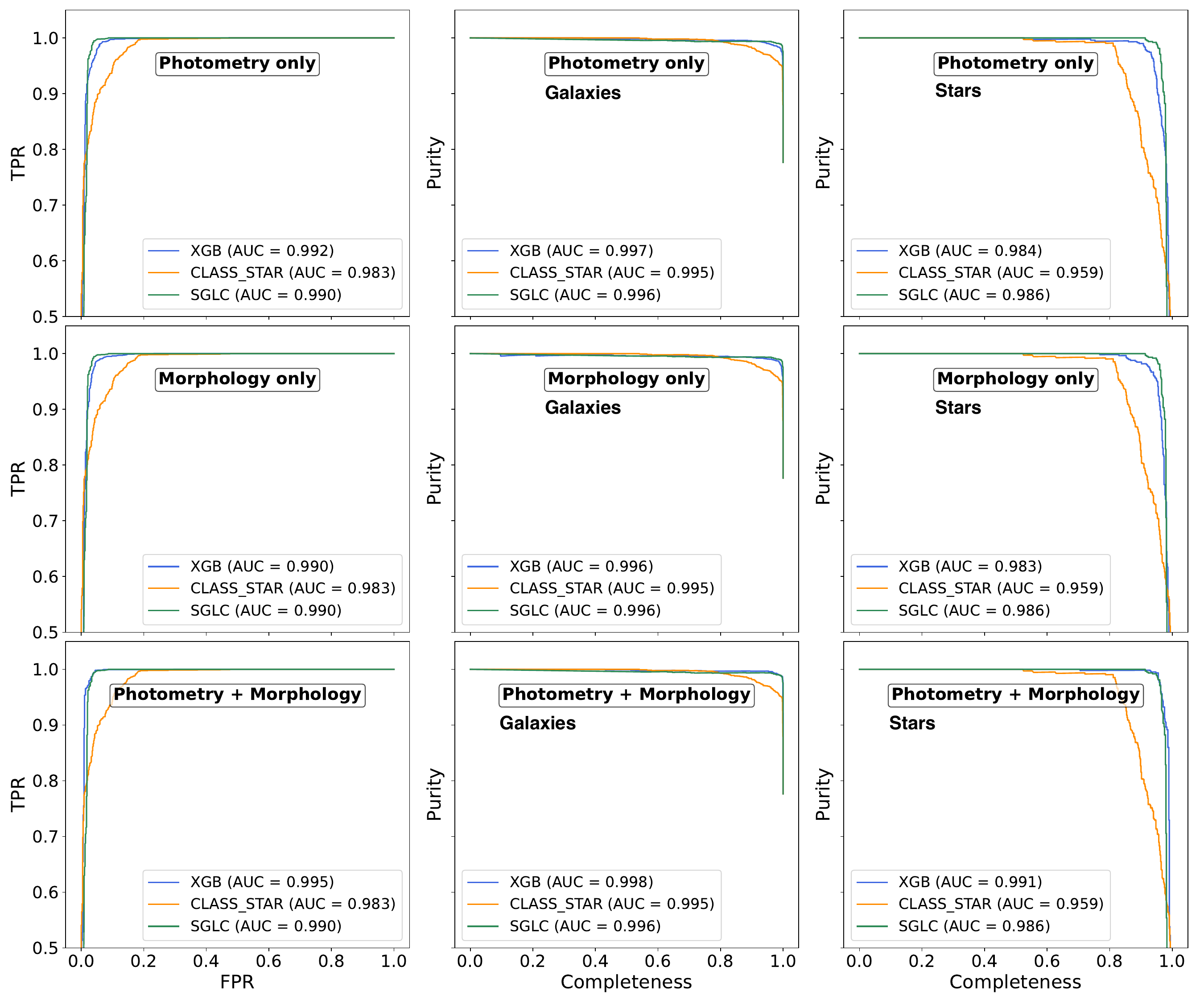}
\caption{ROC and purity–completeness curves for XGB, \mytt{CLASS_STAR}, and SGLC.}\label{fig:rock}
\end{figure}

Figure~\ref{fig:rock} shows the ROC curves for the three XGBoost models: photometry only (123 features, top), morphology only (6 features, middle), and morphology+photometry (128 features, bottom). For comparison, we also include the performance of SGLC and \mytt{CLASS_STAR}, the neural-network classifier implemented in \mytt{SExtractor} \citep{1996A&AS..117..393B}. The photometry-only model outperforms \mytt{CLASS_STAR} and achieves performance comparable to, and slightly better than, SGLC. It also surpasses the morphology-only model, which performs as SGLC. The morphology+photometry model is the best overall, yielding the highest ROC AUC and thus the most accurate classifications.
The same figure also shows the purity and completeness curves for galaxies (second column) and stars (third column). Despite the class imbalance in the training set, stars are well classified, with only a modest reduction in average precision (area under the purity-completeness curve).

\begin{figure}
\centering 
\includegraphics[trim={0 0 0 0}, clip, width= \columnwidth]{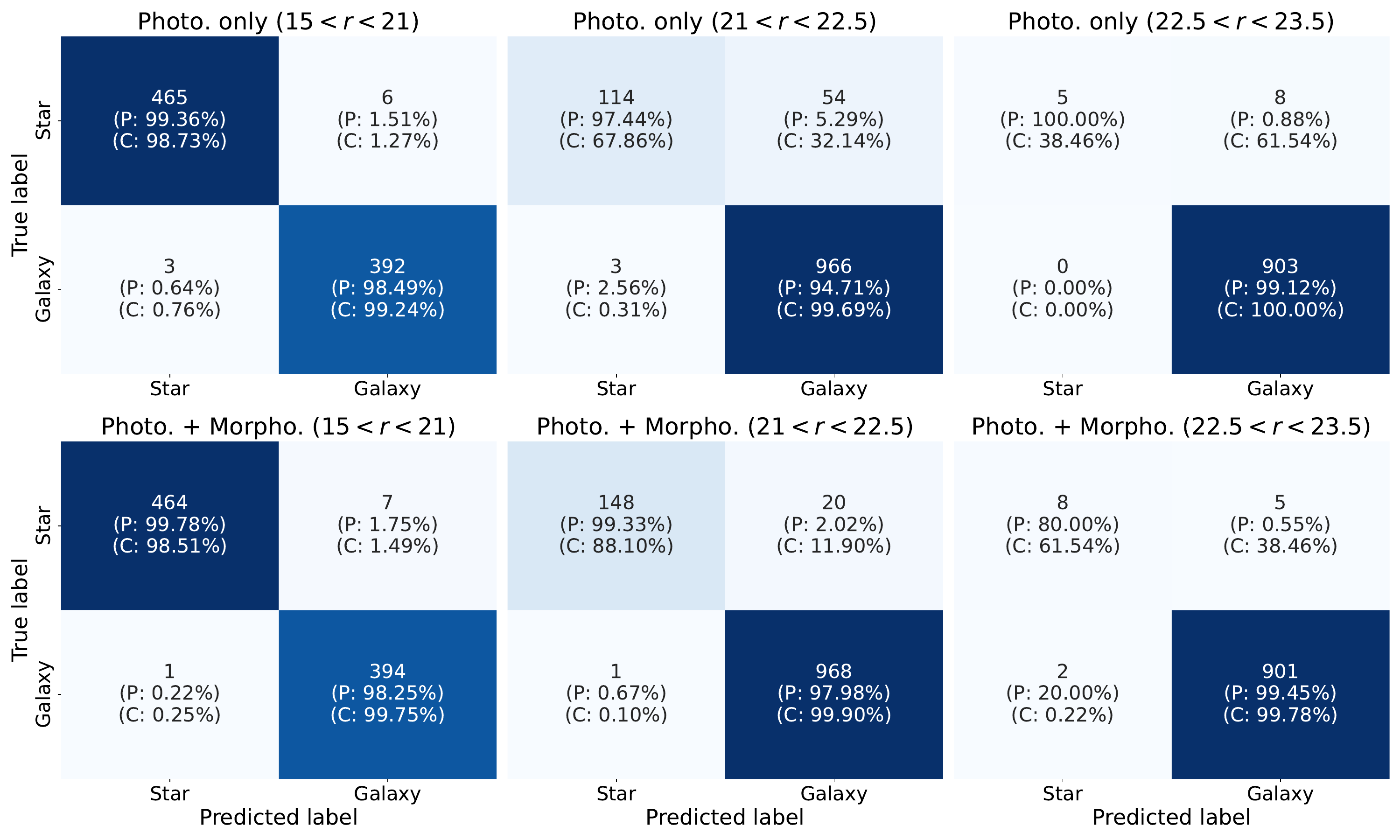}
\caption{Confusion matrices for the XGBoost models using photometry only (top) and photometry plus morphology (bottom), shown for three bins of magnitudes. Each matrix displays the number of objects and associated purity (P) and completeness (C) for each class.}
\label{fig:cm}
\end{figure}

Figure~\ref{fig:cm} shows the confusion matrices for the models based on photometry only (top row) and on photometry combined with morphology (bottom row), evaluated in three $r$-band magnitude bins. A classification threshold of $p_{\rm cut}=0.5$ is used throughout. As expected, classification performance deteriorates at fainter magnitudes, primarily due to stars being increasingly misclassified as galaxies. 
The inclusion of morphological features significantly mitigates this effect, confining most misclassifications to the faintest bin ($r > 22.5$). In contrast, the classification of galaxies remains consistently accurate across all bins.

\begin{figure}
\centering 
\centering 
\includegraphics[trim={0 1.65cm 0 0}, clip, width= \columnwidth]{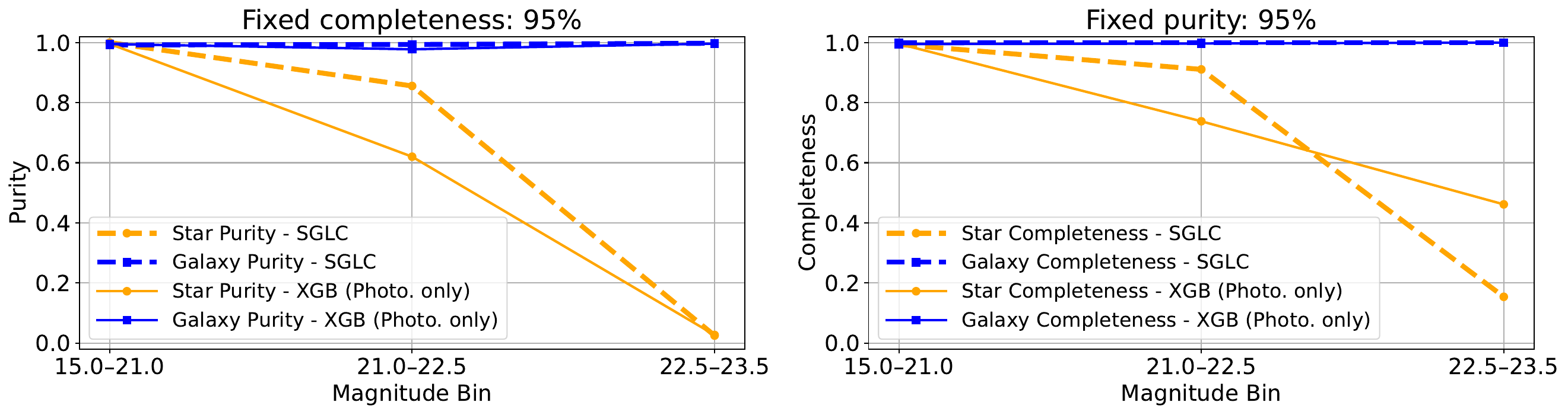}
\includegraphics[trim={0 0 0 1.cm}, clip, width= \columnwidth]{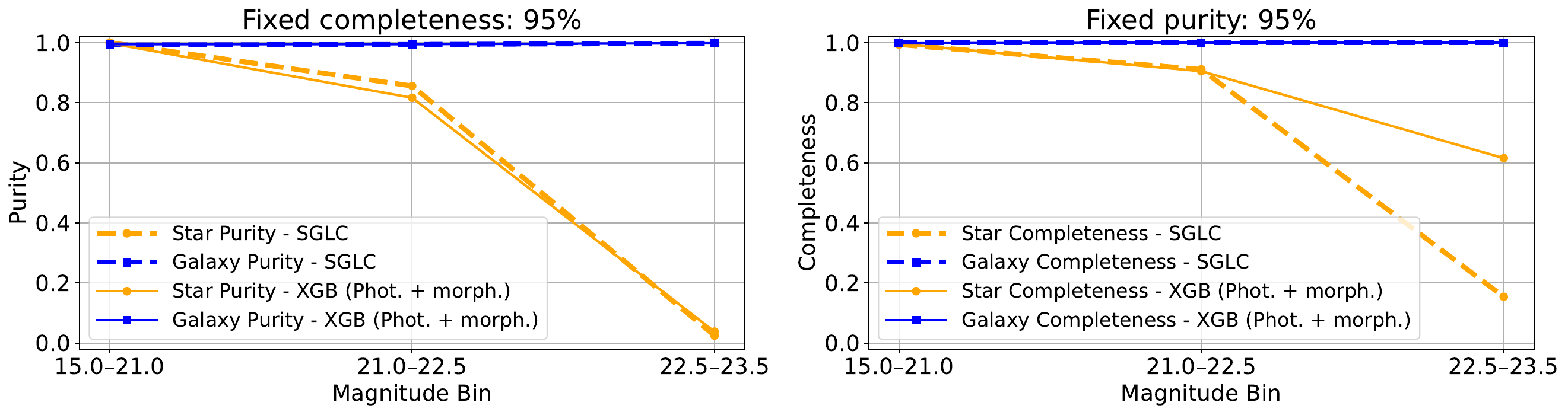}
\includegraphics[trim={0 1.65cm 0 0}, clip, width= \columnwidth]{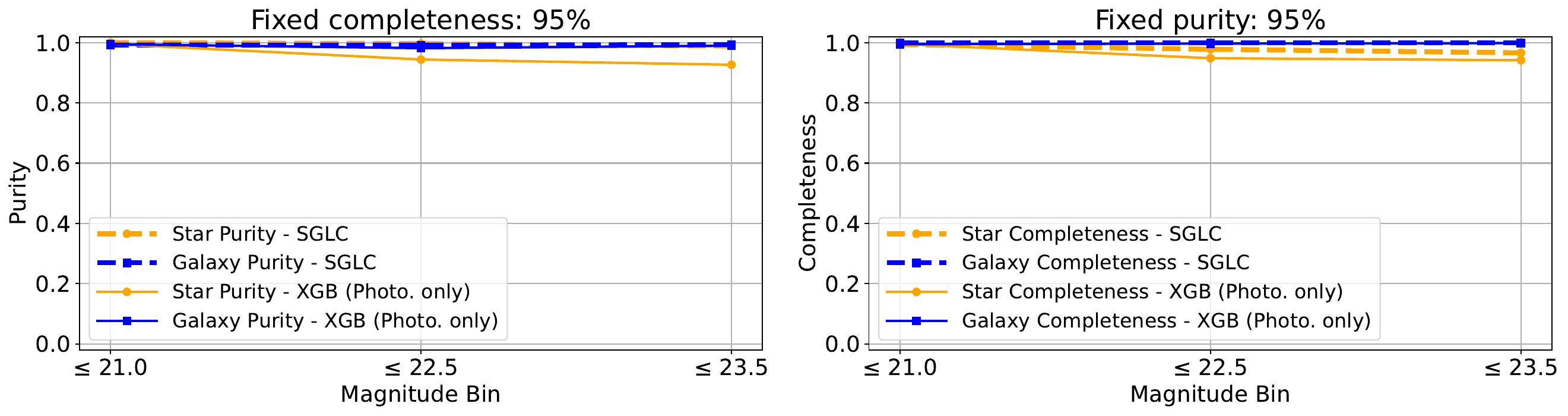}
\includegraphics[trim={0 0 0 1cm}, clip, width= \columnwidth]{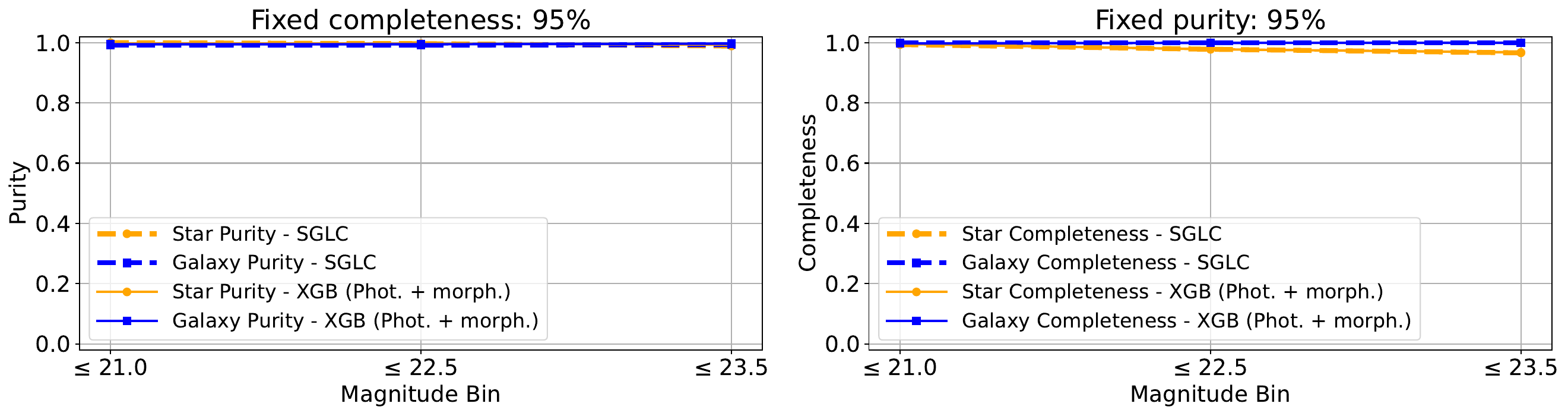}
\caption{Purity (left) and completeness (right) for galaxies and stars as a function of $r$-band magnitude, evaluated at a fixed target value of 95\%. Top panels show results in discrete magnitude bins, while bottom panels display the corresponding cumulative statistics.}
\label{fig:fixpur}
\end{figure}

Figure~\ref{fig:fixpur} shows the purity and completeness for galaxies and stars as functions of the $r$-band magnitude, evaluated at a fixed target value of 95\%. In each magnitude bin, this is achieved by selecting the probability threshold $p_{\rm cut}$ that yields the desired classification level.
Results are presented in discrete magnitude bins (top panels) and cumulatively (bottom panels), for XGBoost models trained with photometry only and with photometry plus morphology, compared against SGLC.  
Galaxy classifications remain highly accurate across all magnitudes. For stars, performance degrades at faint magnitudes, but improves with the inclusion of morphology, surpassing SGLC. The apparent drop at $r > 22.5$ in the binned results is explained by the very small number of faint stars in the training set (see Figure~\ref{fig:jnep-minijpas-mags}) and does not significantly affect the cumulative statistics.

\subsection{Misclassifications}

\begin{figure}
\centering 
\includegraphics[trim={0 0 0 0}, clip, width= .52 \textwidth]{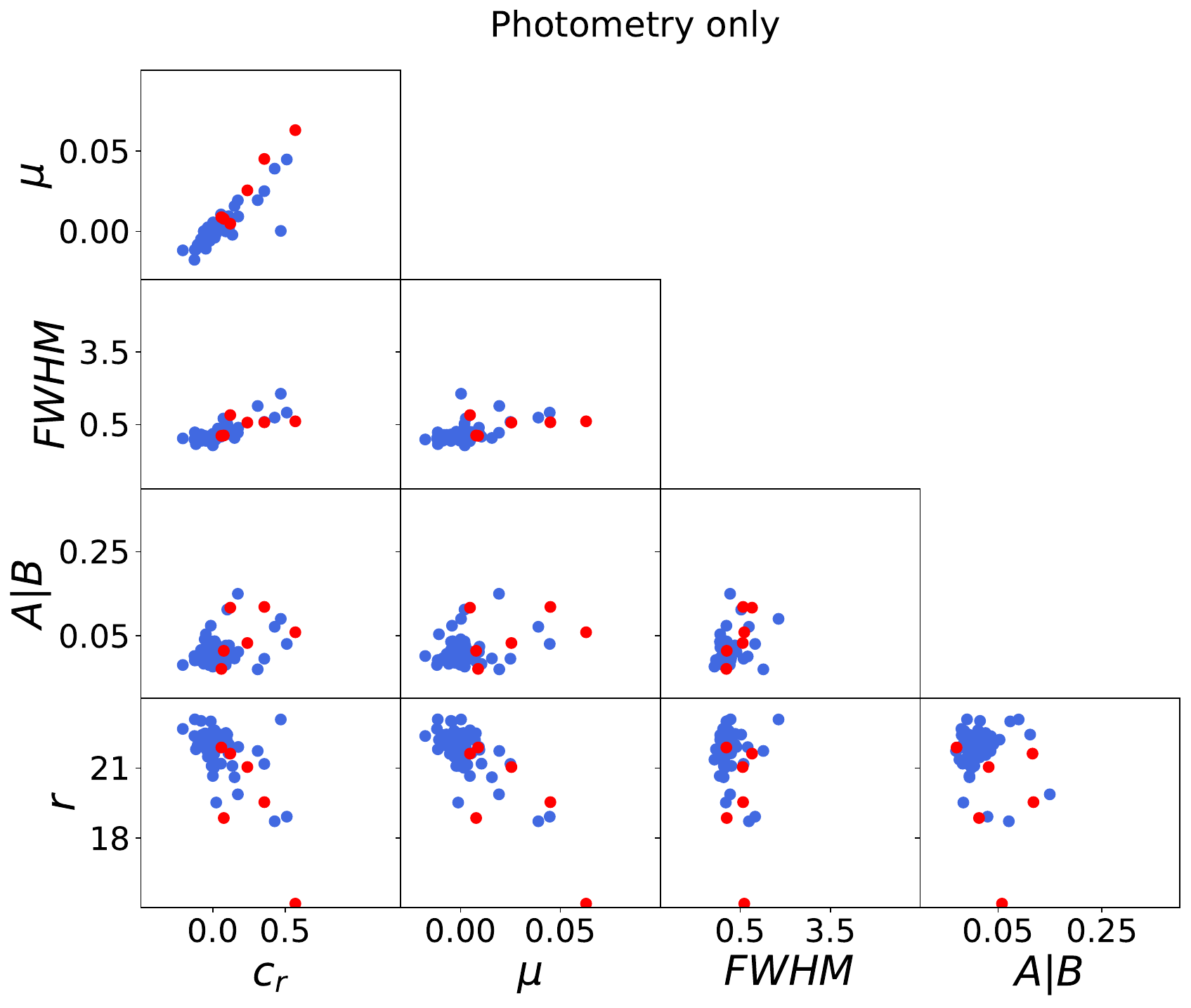}
\includegraphics[trim={0 0 0 0}, clip, width= .465 \textwidth]{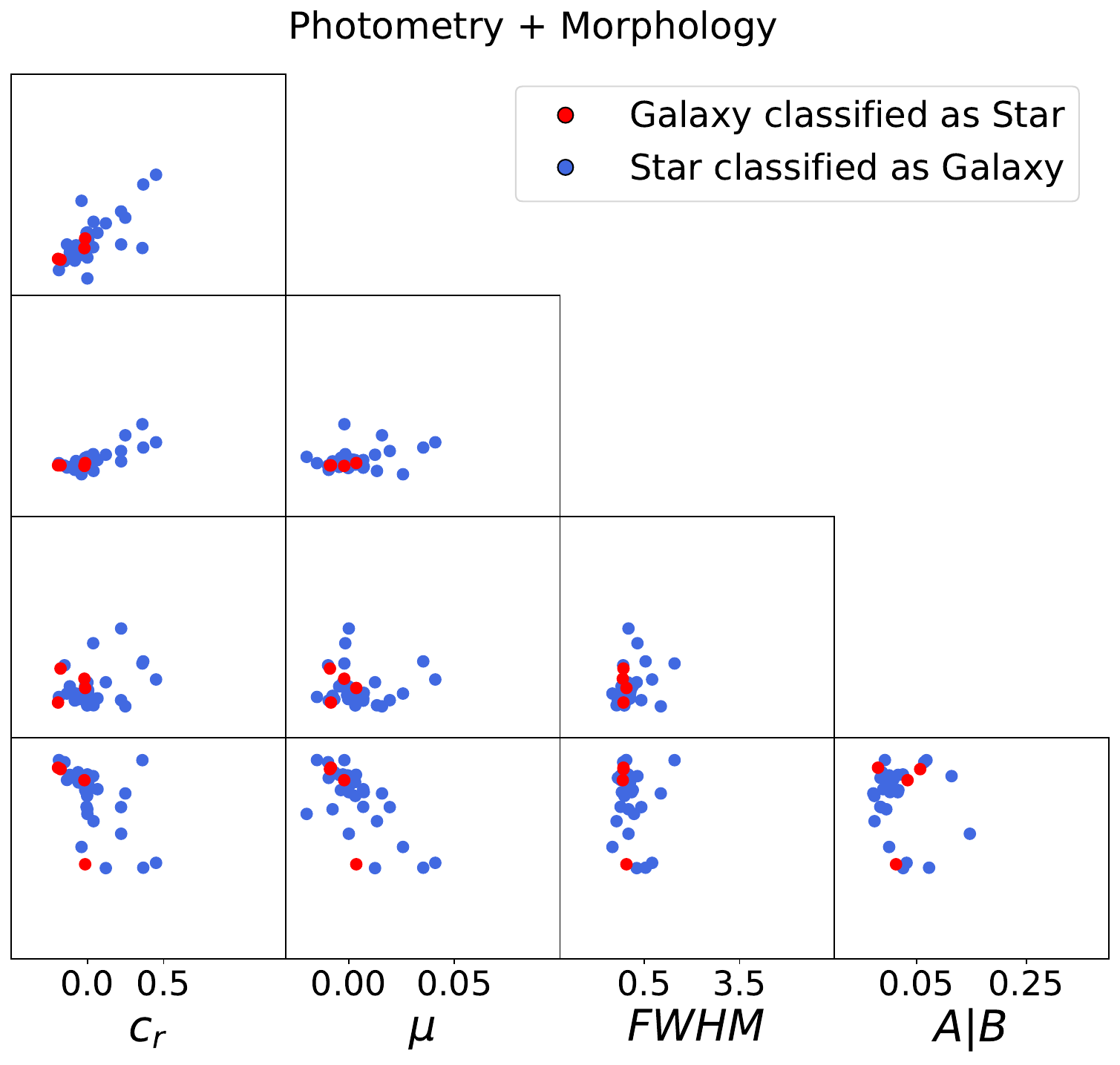}
\caption{Misclassified sources in the normalized morphological parameter space and $r$-band magnitude, for models using photometry only (left) and photometry with morphology (right).}
\label{fig:misses}
\end{figure}

Misclassifications are dominated by faint stars ($r>21$) erroneously labeled as galaxies. Two effects drive this: (i) a brightness imbalance in the training set (relatively brighter stars and fainter galaxies) and (ii) color/SED degeneracy in photometry-only models at low SNR. Adding morphology largely mitigates the problem—star$\rightarrow$galaxy errors drop sharply and galaxy$\rightarrow$star errors are essentially removed. 
At the faint end, however, stellar and galactic morphologies converge, increasing ambiguity. This is evident in Figure~\ref{fig:misses}, where the normalized peak surface brightness $\mu$ drifts toward the stellar reference; the same trend appears in the photometry-only model. Thus, morphology reduces the rate of mistakes but does not fully break the underlying degeneracies, indicating that part of the confusion arises from photometric/SED similarity in faint sources. These results align with \citet{vonMarttens:2022mpv} and suggest that more sophisticated treatment of photometry (e.g., improved uncertainty handling) could yield further gains.

\begin{figure}
\centering 
\includegraphics[trim={0 0 0 0}, clip, width=  \textwidth]{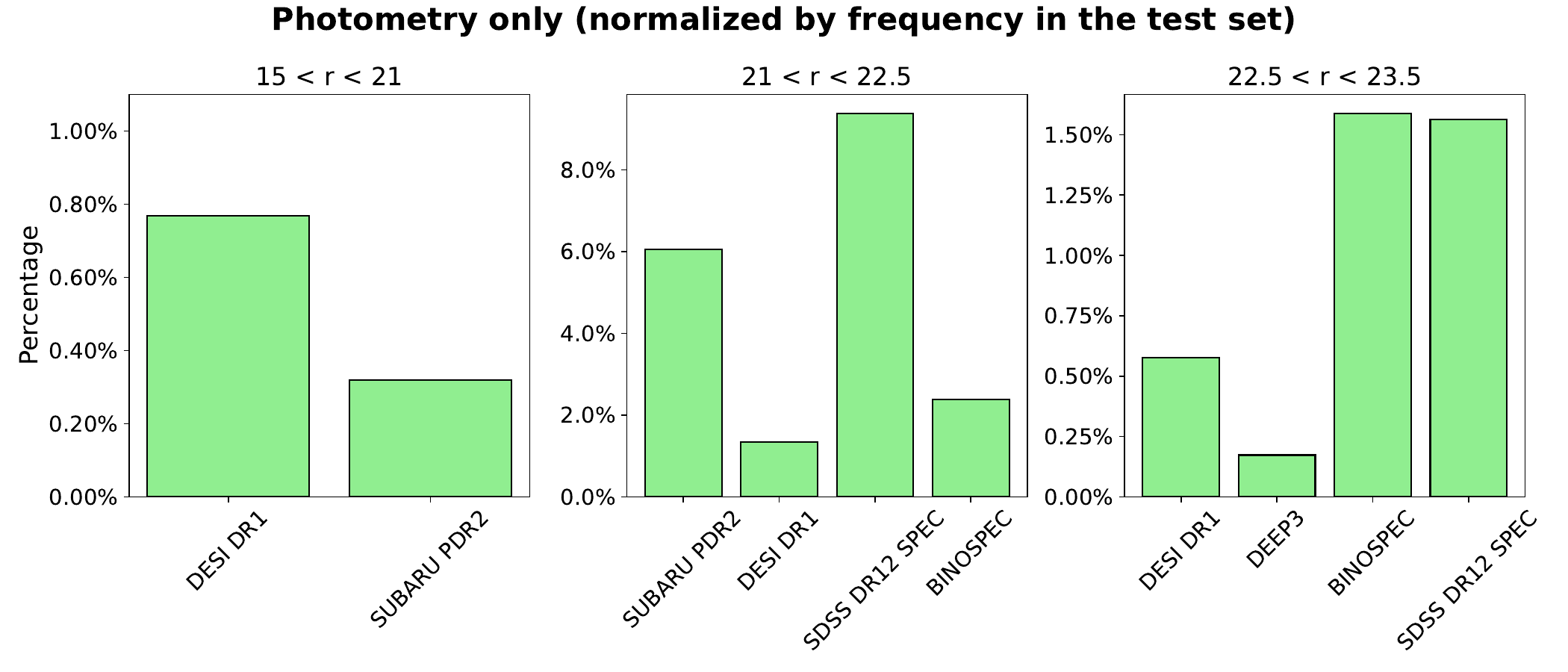}
\includegraphics[trim={0 0 0 0}, clip, width=  \textwidth]{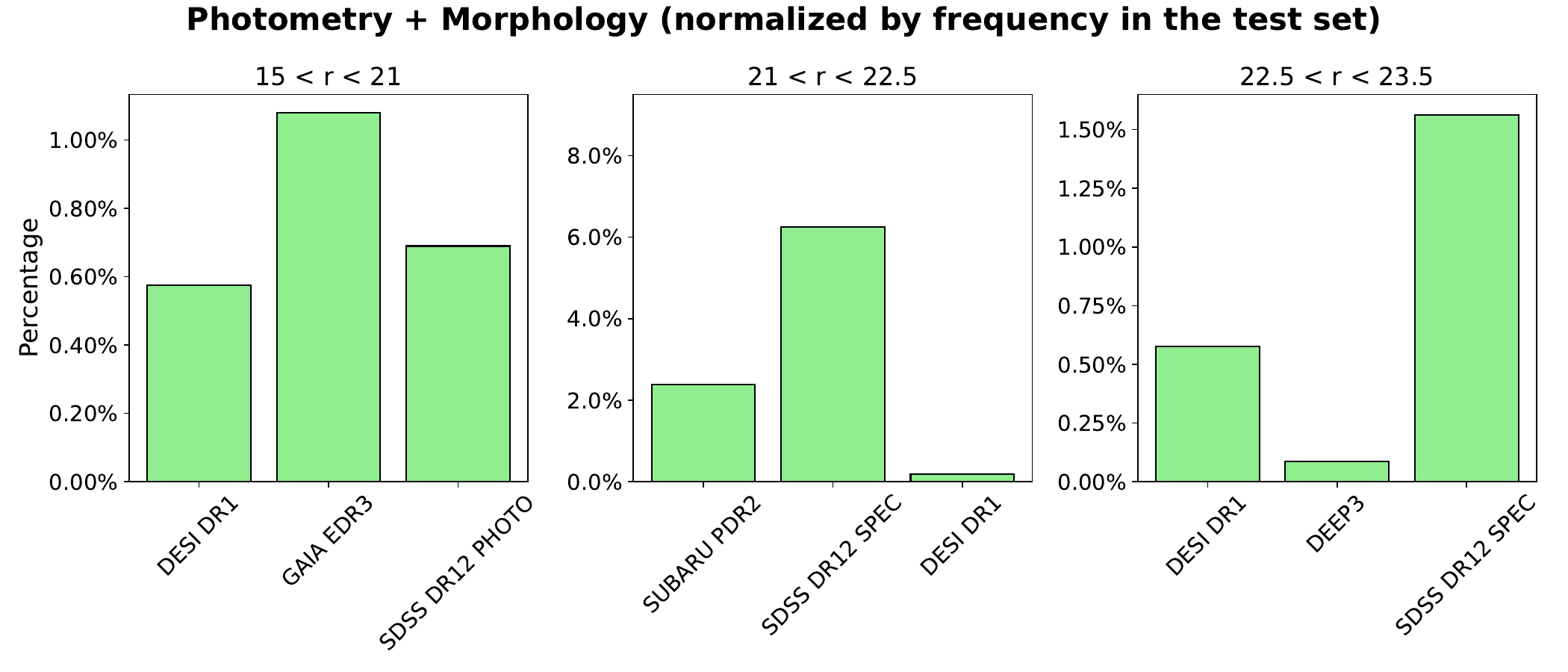}
\caption{Fraction of stars misclassified as galaxies in the test set, normalized by the number of objects per survey, shown in three $r$-band magnitude bins.}
\label{fig:hist-flags-labels}
\end{figure}

To investigate whether additional factors contribute to the misclassifications, 
Figure~\ref{fig:hist-flags-labels} shows the fraction of stars classified as galaxies by survey of origin within the test set. No clear trends are observed, excluding the possibility that stellar misclassifications are mainly due to incorrect truth labels.  
We also examined misclassifications as a function of \mytt{flags}, but the numbers 
were negligible.  
Finally, Appendix~\ref{ap:visual} presents visual inspections of representative 
misclassifications across both models and magnitude ranges.

\subsection{Stellar locus}

\begin{figure}
\centering 
\includegraphics[trim={0 0 0 0}, clip, height= 7.05cm]{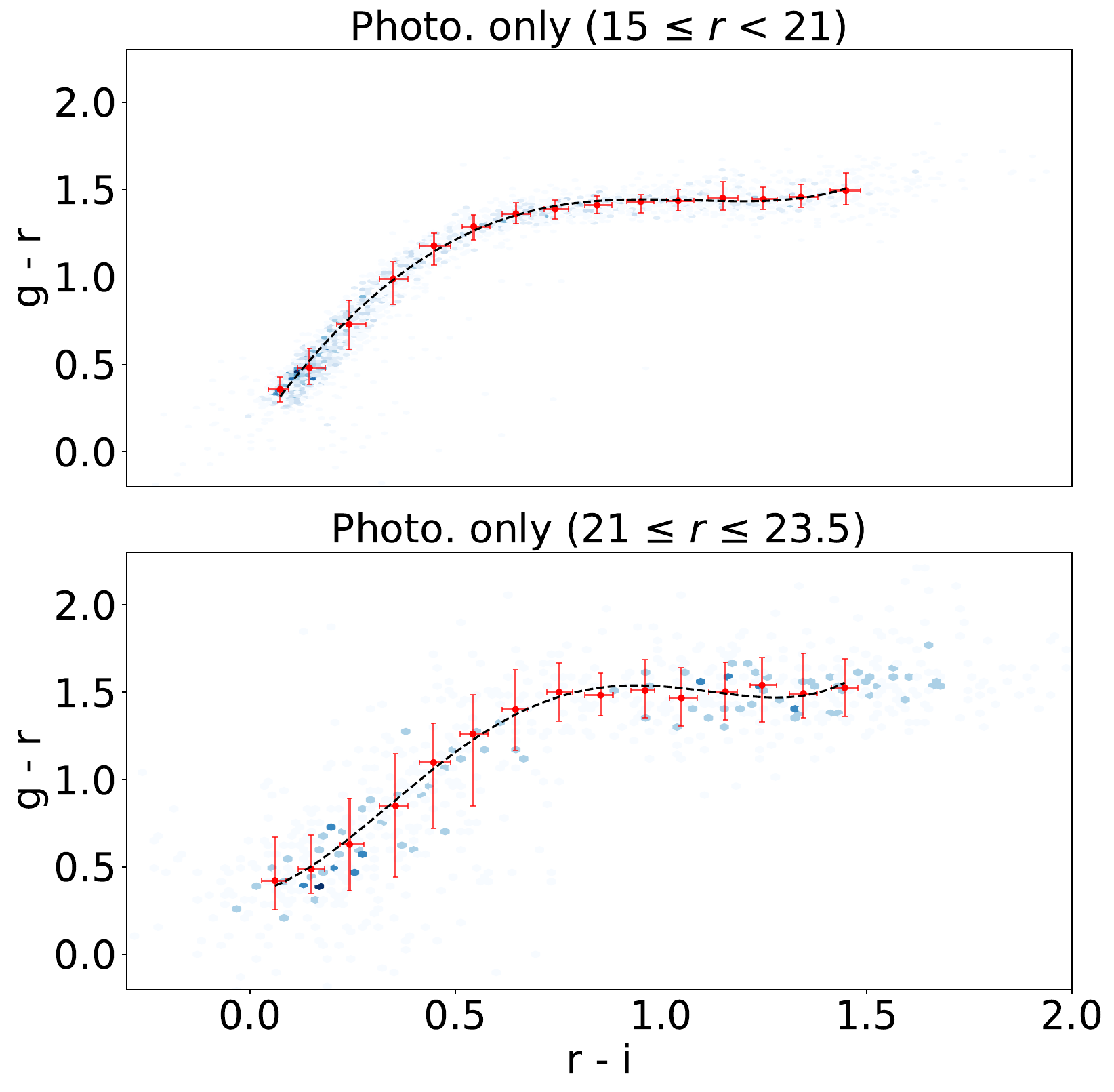}
\includegraphics[trim={0 0 0 0}, clip, height= 7.05cm]{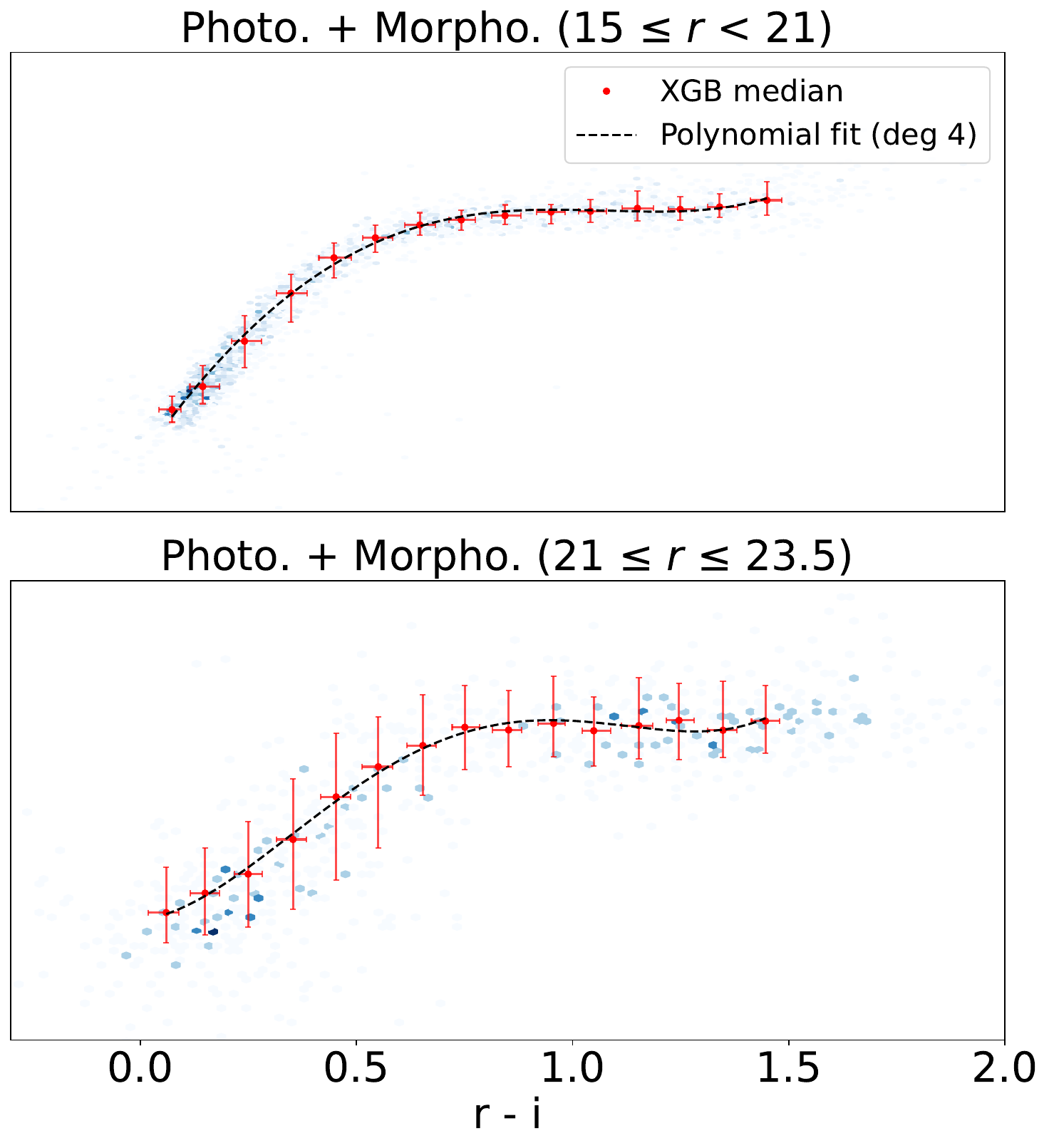}
\caption{Stellar locus: dashed line shows the fit to labeled stars; red markers indicate XGBoost predictions using the model with photometry (left) and photometry and morphology (right) from the full VAC.}
\label{fig:stella-locus}
\end{figure}

Figure~\ref{fig:stella-locus} shows the stellar locus. The dashed line is a fourth-degree polynomial fit to all labeled stars, overplotted on the 2D histogram to guide the reader’s eye. Red markers show the XGBoost predictions from the full value-added catalog (VAC)  using the model with photometry (left) and photometry and morphology (right): each point corresponds to the median value in a bin, with error bars indicating the 16th and 84th percentiles. The predicted locus closely follows the reference across the magnitude range, consistent with the high stellar purity reported in Figure~\ref{fig:cm}.

\subsection{Feature importance}

To evaluate the relative contribution of each feature group, we computed permutation
importance on the test set. This metric measures the change in ROC AUC when the values
of a given feature group are randomly permuted, thereby breaking its relationship with
the target. A larger drop corresponds to higher importance.
Small negative values (i.e., a slight increase in AUC after permutation) do not imply
that a feature 'confuses' the classifier; they typically arise from sampling noise on a
finite test set and/or correlations with other features. We therefore interpret such
negative importances as statistically consistent with zero (non-informative) importance.

\begin{figure}
\centering 
\includegraphics[trim={0 0 0 0}, clip, width= \columnwidth]{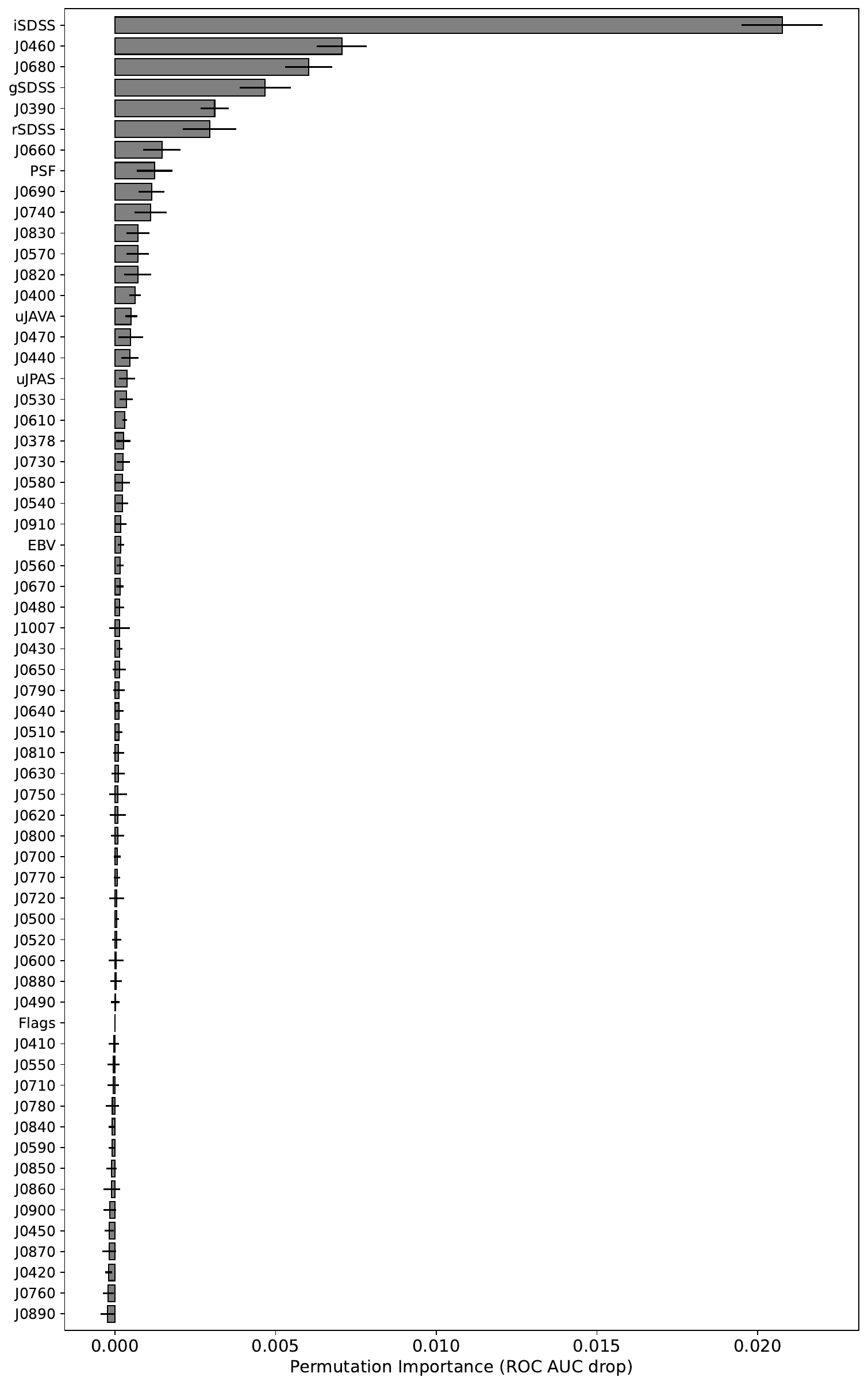}
\caption{Permutation importance, measured as the decrease in ROC AUC when each group is shuffled, using the XGBoost model trained with photometry only. Each feature group includes both its value and its associated error.}
\label{fig:fi1}
\end{figure}

Figure~\ref{fig:fi1} shows the permutation importance results for the XGBoost model trained with photometry only.
The most informative features are the three broad bands $i$, $g$, and $r$, together with three narrow bands: J0460 (4603~\AA), J0680 (6812~\AA), and J0390 (3904~\AA). These results are consistent with those of \citet[][Figure 13]{Baqui:2020sfd}, who found peaks in feature importance around 4000 and 6900~\AA. This suggests that the model exploits these wavelengths to estimate the slope of the spectral energy distribution.

\begin{figure}
\centering 
\includegraphics[trim={0 0 0 0}, clip, width= \columnwidth]{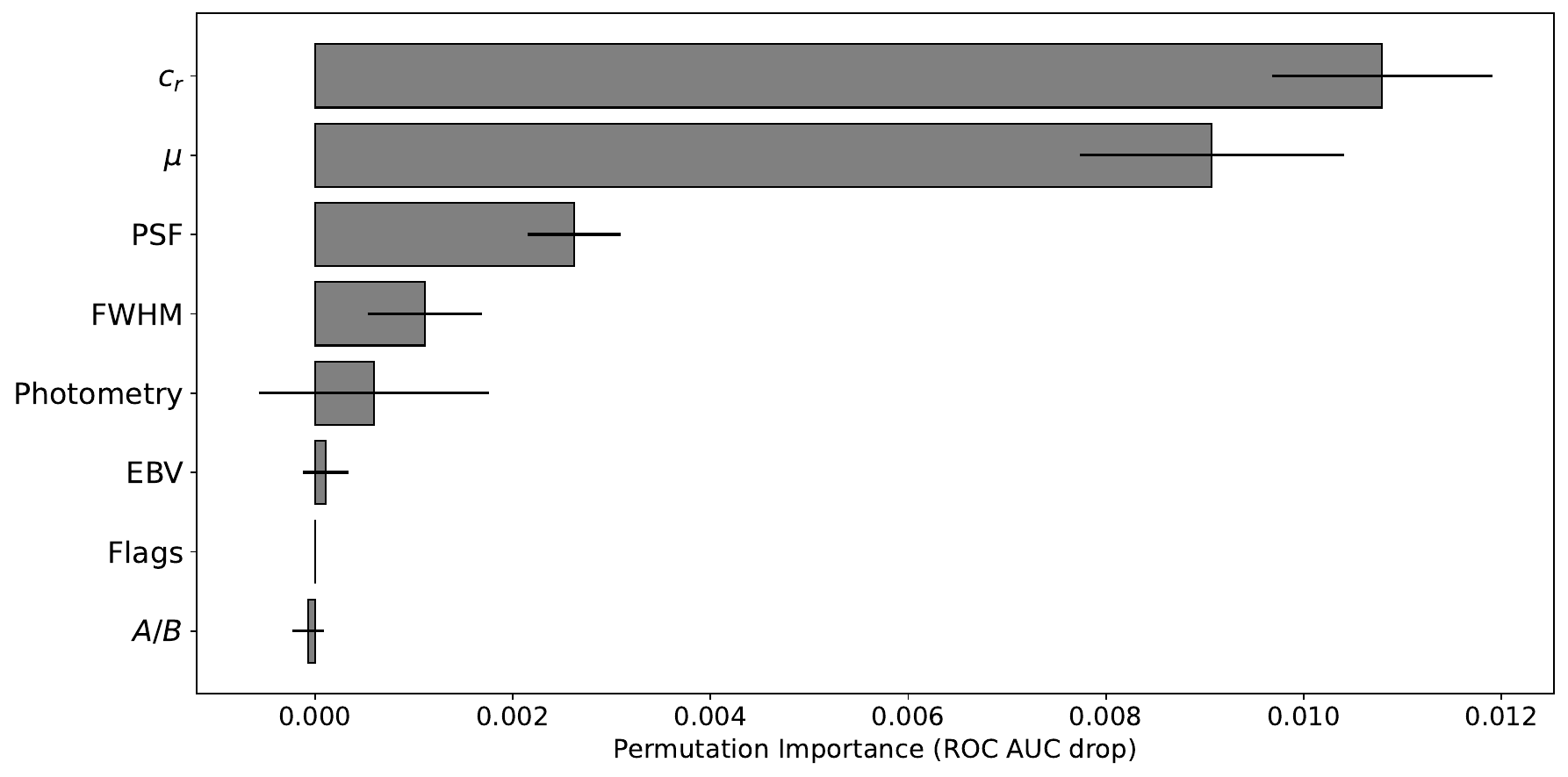}
\caption{Permutation importance of feature groups, measured as the decrease in ROC AUC when each group is shuffled, using the XGBoost model trained with photometry and morphology. Photometry refers to the combined importance of all 60 bands and their associated errors; EBV includes both the extinction value and its error.}
\label{fig:fi2}
\end{figure}

Figure~\ref{fig:fi2} shows the results for the XGBoost model trained with photometry and morphology. The most informative features are  the concentration index $c_r$ and the normalized peak surface brightness $\mu$, followed by the PSF. FWHM and photometry also contribute, while ellipticity $(A/B)$, extinction $E(B-V)$ with its error, and detection flags have negligible impact. 
This reflects the fact that miniJPAS and J-NEP each cover only about $1\deg^2$, where Milky Way extinction varies very little. The results therefore confirm that morphological features carry the dominant information for effective star–galaxy separation in these fields.

\section{Conclusions}
\label{sec:conclu}

The machine learning-based classification of J-NEP and miniJPAS sources presented in this paper improves star-galaxy separation compared to previous methods. Using an XGBoost classifier with hyperparameters tuned through automated optimization, trained on a robust labeled dataset derived from crossmatching with multiple spectroscopic and photometric catalogs, we provide reliable classifications both from purely photometric data and from combined photometric and morphological features.

Our key findings can be summarized as follows:
\begin{enumerate}

\item The XGBoost model combining photometric and morphological features achieves a higher area under the ROC curve and outperforms both the SGLC and \mytt{CLASS_STAR} classifiers across all magnitude ranges. 

\item Using photometry alone yields competitive performance, slightly surpassing SGLC, but leads to  misclassifications of faint stars as galaxies at $r > 22$. This ambiguity is  reduced when morphological features are included.

\item The permutation importance analysis identifies the concentration, normalized peak surface brightness, and PSF as critical morphological parameters for classification. Photometric bands at approximately 3900, 4600, and 6800~\AA\ are consistently among the most informative, highlighting their relevance for characterizing stellar and galactic spectral slopes.

\item The UMAP representativeness analysis demonstrates the robustness and consistency of the labeled training set, validating its applicability for reliable classification across the full miniJPAS and J-NEP datasets.

\item Our ML pipeline, publicly released as a value-added catalog (VAC), offers reliable star-galaxy classifications that can directly support downstream scientific analyses, including studies of galaxy evolution and cosmological investigations reliant on precise object classifications.

\end{enumerate}

\vspace{6pt}

\authorcontributions{
Conceptualization, VM;
methodology, APJ, GVS, VM and RvM;
software, APJ and GVS;
validation, APJ, GVS, VM and RvM;
formal analysis, APJ, GVS, VM and RvM;
investigation, APJ, GVS, VM and RvM;
resources, R.A., J.A., N.B., S.B., J.C., D.C.H., S.D., R.D., A.E., R.M.G.D., A.H.C., C.H.M., J.L., C.L.S., A.M.F., C.M.O., M.M., F.R., L.S., K.T., J.V., H.V.R., J.M.V.,  C.W., J.Z.C.; 
data curation, APJ, GVS and VM;
writing---original draft preparation, VM;
writing---review and editing, APJ, GVS, VM, RvM, SGL, CW;
visualization, APJ and GVS;
supervision, VM;
project administration, VM;
funding acquisition, V.M., R.A., J.A., N.B., S.B., J.C., D.C.H., S.D., R.D., A.E., R.M.G.D., A.H.C., C.H.M., J.L., C.L.S., A.M.F., C.M.O., M.M., F.R., L.S., K.T., J.V., H.V.R., J.M.V., J.Z.C.
All authors have read and agreed to the published version of the manuscript.}


\funding{
APJ acknowledges financial support from CAPES (Brazil).
GVS acknowledges financial support from UFES (Brazil).
VM acknowledges partial support from CNPq (Brazil) and FAPES (Brazil).
RvM is suported by Fundação de Amparo à Pesquisa do Estado da Bahia (FAPESB) grant TO APP0039/2023.
SGL acknowledge the financial support from the MICIU with funding from the European Union NextGenerationEU and Generalitat Valenciana in the call Programa de Planes Complementarios de I+D+i (PRTR 2022) Project (VAL-JPAS), reference ASFAE/2022/025. SGL  also acknowledge  the research Project PID2023-149420NB-I00 funded by MICIU/AEI/10.13039/501100011033 and by ERDF/EU and the project of excellence PROMETEO CIPROM/2023/21 of the Conselleria de Educación, Universidades y Empleo (Generalitat Valenciana).
RGD acknowledges financial support from the Severo Ochoa grant CEX2021-001131-S, funded by MICIU/AEI (10.13039/501100011033), and is also grateful for support from project PID2022-141755NB-I00.}

\dataavailability{
The data and code underlying this article are available from the corresponding author upon reasonable request. 
The value-added catalog will be publicly released in accordance with J-PAS collaboration policies, through the tables \mytt{jnep.StarGalClass} and \mytt{minijpas.StarGalClass} at \href{https://www.j-pas.org/datareleases}{j-pas.org/datareleases}.
}

\acknowledgments{
This paper has undergone internal review by the J-PAS Collaboration.\\
This work made use of computational resources provided by the  \href{https://computacaocientifica.ufes.br/scicom}{Sci-Com Lab} of the Department of Physics at UFES, supported by FAPES, CAPES and CNPq.\\
Based on observations made with the JST/T250 telescope and JPCam at the Observatorio
Astrofísico de Javalambre (OAJ), in Teruel, owned, managed, and operated by the Centro de
Estudios de Física del Cosmos de Aragón (CEFCA). We acknowledge the OAJ Data Processing and
Archiving Unit (UPAD) for reducing and calibrating the OAJ data used in this work.\\
Funding for the J-PAS Project has been provided by the Governments of Spain and Arag\'on through the Fondo de Inversiones de Teruel; the Aragonese Government through the Research Groups E96, E103, E16\_17R, E16\_20R, and E16\_23R; the Spanish Ministry of Science and Innovation (MCIN/AEI/10.13039/501100011033 y FEDER, Una manera de hacer Europa) with grants PID2021-124918NB-C41, PID2021-124918NB-C42, PID2021-124918NA-C43, and PID2021-124918NB-C44; the Spanish Ministry of Science, Innovation and Universities (MCIU/AEI/FEDER, UE) with grants PGC2018-097585-B-C21 and PGC2018-097585-B-C22; the Spanish Ministry of Economy and Competitiveness (MINECO) under AYA2015-66211-C2-1-P, AYA2015-66211-C2-2, and AYA2012-30789; and European FEDER funding (FCDD10-4E-867, FCDD13-4E-2685). The Brazilian agencies FINEP, FAPESP, FAPERJ and the National Observatory of Brazil have also contributed to this project. Additional funding was provided by the Tartu Observatory and by the J-PAS Chinese Astronomical Consortium.
}

\conflictsofinterest{The authors declare no conflicts of interest.}



\appendixtitles{yes} 
\appendixstart

\appendix

\section[\appendixname~\thesection]{Queries}\label{ap:queries}

\subsection[\appendixname~\thesubsection]{miniJPAS and J-NEP sources}

\noindent
Query used to select miniJPAS sources (an analogous query is used for J-NEP):
\begin{verbatim}
SELECT
    t1.NUMBER, t1.TILE_ID,
FROM
    minijpas.MagABDualObj t1
WHERE
    t1.flags[minijpas::rSDSS] IN (0,1,2,4,6)
    AND t1.mask_flags[minijpas::rSDSS] = 0
    AND t1.MAG_AUTO[minijpas::rSDSS] < 23.5
    AND t1.MAG_AUTO[minijpas::rSDSS] > 15
\end{verbatim}

\subsection[\appendixname~\thesubsection]{SDSS DR12 photometric crossmatch}

\noindent
Query used to extract miniJPAS sources with SDSS DR12 crossmatch and reliable photometric classification (an analogous query is used for J-NEP):
\begin{verbatim}
SELECT
    t1.NUMBER, t1.TILE_ID,
    t2.class AS label
FROM
    minijpas.MagABDualObj t1
JOIN
    minijpas.xmatch_sdss_dr12 t2
    ON t1.TILE_ID = t2.TILE_ID AND t1.NUMBER = t2.NUMBER
WHERE
    t1.flags[minijpas::rSDSS] IN (0,1,2,4,6)
    AND t1.mask_flags[minijpas::rSDSS] = 0
    AND t1.MAG_AUTO[minijpas::rSDSS] < 20
    AND t1.MAG_AUTO[minijpas::rSDSS] > 15
\end{verbatim}

\subsection[\appendixname~\thesubsection]{SDSS DR12 spectroscopic crossmatch}

\noindent
Query used for miniJPAS sources with SDSS  spectroscopic crossmatch for $r > 20$:
\begin{verbatim}
SELECT
    t1.NUMBER, t1.TILE_ID,
    t2.spCl AS label
FROM
    minijpas.MagABDualObj t1
JOIN
    minijpas.xmatch_sdss_dr12 t2
    ON t1.TILE_ID = t2.TILE_ID AND t1.NUMBER = t2.NUMBER
WHERE
    t1.flags[minijpas::rSDSS] IN (0,1,2,4,6)
    AND t1.mask_flags[minijpas::rSDSS] = 0
    AND t1.MAG_AUTO[minijpas::rSDSS] < 23.5
    AND t1.MAG_AUTO[minijpas::rSDSS] > 20
    AND t2.f_zsp = 0
\end{verbatim}

\subsection[\appendixname~\thesubsection]{Gaia EDR3 crossmatch}

\noindent
Query used to extract miniJPAS stars via Gaia parallax (also for J-NEP):
\begin{verbatim}
SELECT
    t1.NUMBER, t1.TILE_ID,
    t2.parallax_over_error AS label
FROM
    minijpas.MagABDualObj t1
JOIN
    minijpas.xmatch_gaia_edr3 t2
    ON t1.TILE_ID = t2.TILE_ID AND t1.NUMBER = t2.NUMBER
WHERE
    t1.flags[minijpas::rSDSS] IN (0,1,2,4,6)
    AND t1.mask_flags[minijpas::rSDSS] = 0
    AND t1.MAG_AUTO[minijpas::rSDSS] < 23.5
    AND t1.MAG_AUTO[minijpas::rSDSS] > 15
    AND t2.parallax_over_error > 3
\end{verbatim}

\subsection[\appendixname~\thesubsection]{HSC PDR2 crossmatch}

\noindent
Query used to extract miniJPAS sources crossmatched with HSC PDR2:
\begin{verbatim}
SELECT
    t1.NUMBER, t1.TILE_ID,
    t2.g_extendedness_value AS label_g,
    t2.r_extendedness_value AS label_r,
    t2.i_extendedness_value AS label_i,
    t2.z_extendedness_value AS label_z,
    t2.y_extendedness_value AS label_y
FROM
    minijpas.MagABDualObj t1
JOIN
    minijpas.xmatch_subaru_pdr2 t2
    ON t1.TILE_ID = t2.TILE_ID AND t1.NUMBER = t2.NUMBER
WHERE
    t1.flags[minijpas::rSDSS] IN (0,1,2,4,6)
    AND t1.mask_flags[minijpas::rSDSS] = 0
    AND t1.MAG_AUTO[minijpas::rSDSS] < 23.5
    AND t1.MAG_AUTO[minijpas::rSDSS] > 18.5
    AND t2.g_inputcount_value >= 4
    AND t2.g_cmodel_flag = false
    AND t2.g_pixelflags = false
    AND t2.r_inputcount_value >= 4
    AND t2.r_cmodel_flag = false
    AND t2.r_pixelflags = false
    AND t2.i_inputcount_value >= 4
    AND t2.i_cmodel_flag = false
    AND t2.i_pixelflags = false
    AND t2.z_inputcount_value >= 4
    AND t2.z_cmodel_flag = false
    AND t2.z_pixelflags = false
    AND t2.y_inputcount_value >= 4
    AND t2.y_cmodel_flag = false
    AND t2.y_pixelflags = false
\end{verbatim}

\section[\appendixname~\thesection]{Hyperparameter configuration} \label{ap:hp}

The best-performing pipeline identified by TPOT consists of a \mytt{MaxAbsScaler} for feature scaling followed by an \mytt{XGBClassifier}. The choice of XGBoost is further supported by \citet{vonMarttens:2022mpv}, which--although based on 12 photometric bands in J-PLUS rather than the 60 in J-PAS--uses the same morphological features and comparable photometric coverage. In that study, TPOT systematically evaluated multiple machine learning algorithms and identified XGBoost as the top performer. The consistency between the two surveys thus provides a strong justification for adopting XGBoost in our analysis.
The final hyperparameter configuration is as follows:
\begin{itemize}
    \item \mytt{booster = gbtree}: specifies the tree-based boosting model, suitable for structured tabular data.
    \item \mytt{colsample_bylevel = 0.6}: fraction of features sampled for each tree level, reducing overfitting by introducing randomness.
    \item \mytt{colsample_bynode = 1.0}: fraction of features sampled for each split, set to 1 (all features used).
    \item \mytt{colsample_bytree = 0.9}: fraction of features sampled for each tree, promoting diversity among trees.
    \item \mytt{gamma = 0.5}: minimum loss reduction required for further partitioning; larger values make the model more conservative.
    \item \mytt{learning_rate = 0.05}: step size shrinkage, balancing learning speed with generalization.
    \item \mytt{max_depth = 7}: maximum depth of each tree, controlling model complexity.
    \item \mytt{min_child_weight = 5}: minimum sum of instance weights in a child node; higher values prevent overfitting small fluctuations.
    \item \mytt{n_estimators = 600}: number of boosting rounds (trees).
    \item \mytt{objective = binary:logistic}: specifies binary classification with probabilistic outputs.
    \item \mytt{reg_alpha = 0}: L1 regularization term, promoting sparsity in weights; set to zero here.
    \item \mytt{reg_lambda = 1}: L2 regularization term, penalizing large weights to reduce overfitting.
    \item \mytt{scale_pos_weight = 3.46}: weighting factor to address class imbalance between stars and galaxies.
    \item \mytt{subsample = 0.95}: fraction of training instances sampled for each tree, preventing overfitting by adding stochasticity.
\end{itemize}
This configuration balances model flexibility and regularization, effectively handling the imbalance between stellar and galactic classes while maintaining robustness against overfitting.

\section[\appendixname~\thesection]{Treatment of Ambiguous Crossmatches} 
\label{ap:ambiguous}
\renewcommand{\thefigure}{C\arabic{figure}}
\renewcommand{\thetable}{C\arabic{table}}
\setcounter{figure}{0}
\setcounter{table}{0}

\begin{figure}
\raggedleft 
\includegraphics[trim={0 1.35cm 0 0.35cm}, clip, width= .95\columnwidth]{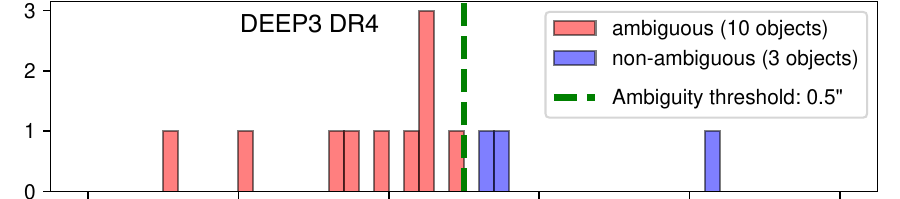}
\includegraphics[trim={0 1.35cm 0 0.35cm}, clip, width= .965\columnwidth]{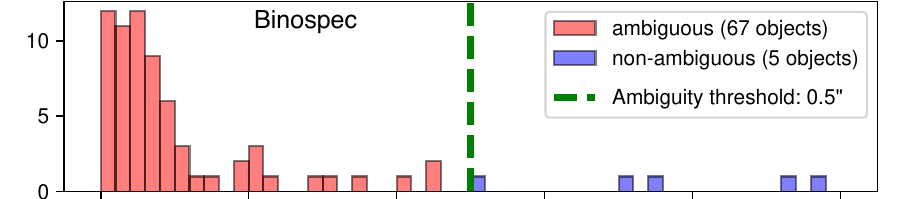}
\includegraphics[trim={0 0 0 0.35cm}, clip, width= \columnwidth]{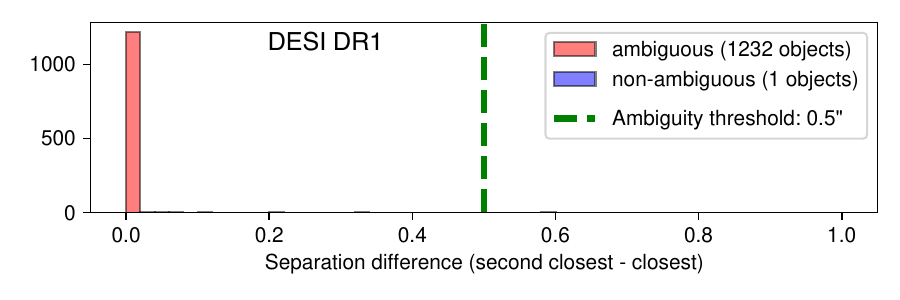}
\caption{Distribution of ambiguous and non-ambiguous crossmatches. The horizontal axis shows the separation difference between the closest and second-closest sources matched to the same miniJPAS or J-NEP object. Red bars indicate ambiguous matches (within 0.5 arcsec), and blue bars show non-ambiguous cases.}
\label{fig:ambiguous}
\end{figure}

In crossmatching miniJPAS and J-NEP with external spectroscopic catalogs (DEEP3, Binospec, and DESI), ambiguous matches may arise when multiple J-PAS sources fall within the matching radius of the same external catalog object. 
By default, \texttt{onexmatch} resolves duplicates by retaining only the closest J-PAS source for each catalog entry. However, when two candidate sources are nearly equidistant from the same object, this procedure may not be reliable. To address such cases, we adopted an ambiguity threshold of 0.5 arcsec, which is approximately twice the pixel scale and comparable to the PSF size. If the positional difference between two candidate matches is below this threshold, the association is deemed ambiguous and the sources are discarded from the final crossmatch, see Figure~\ref{fig:ambiguous}.
This conservative approach minimizes spurious associations, ensuring that the resulting training set retains only high-confidence matches.

\section[\appendixname~\thesection]{Visual inspection of misclassifications} 
\label{ap:visual}
\renewcommand{\thefigure}{D\arabic{figure}}
\renewcommand{\thetable}{D\arabic{table}}
\setcounter{figure}{0}
\setcounter{table}{0}

Here we present a visual inspection of misclassifications produced by the models trained with photometric information only and with combined photometric and morphological information. Representative examples were selected to cover the full magnitude range considered in this work. This qualitative analysis complements the statistical results by illustrating typical failure modes and offering insights into possible improvements for future modeling.

\begin{figure}[p]
\centering 
\includegraphics[trim={0 0 0 0}, clip, height=11cm, width= .49\linewidth]{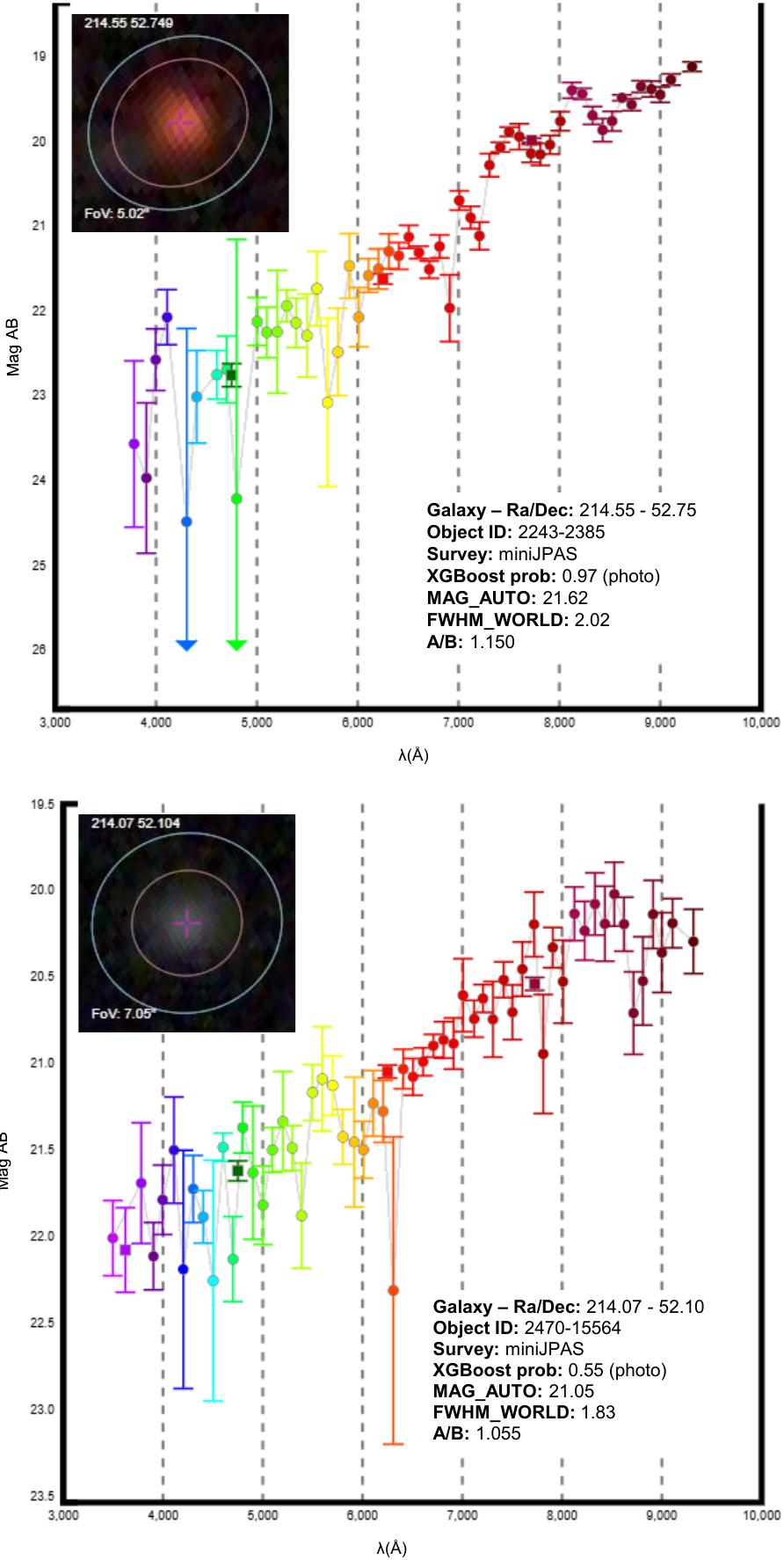}
\includegraphics[trim={0 0 0 0}, clip, height=11cm, width= .49\linewidth]{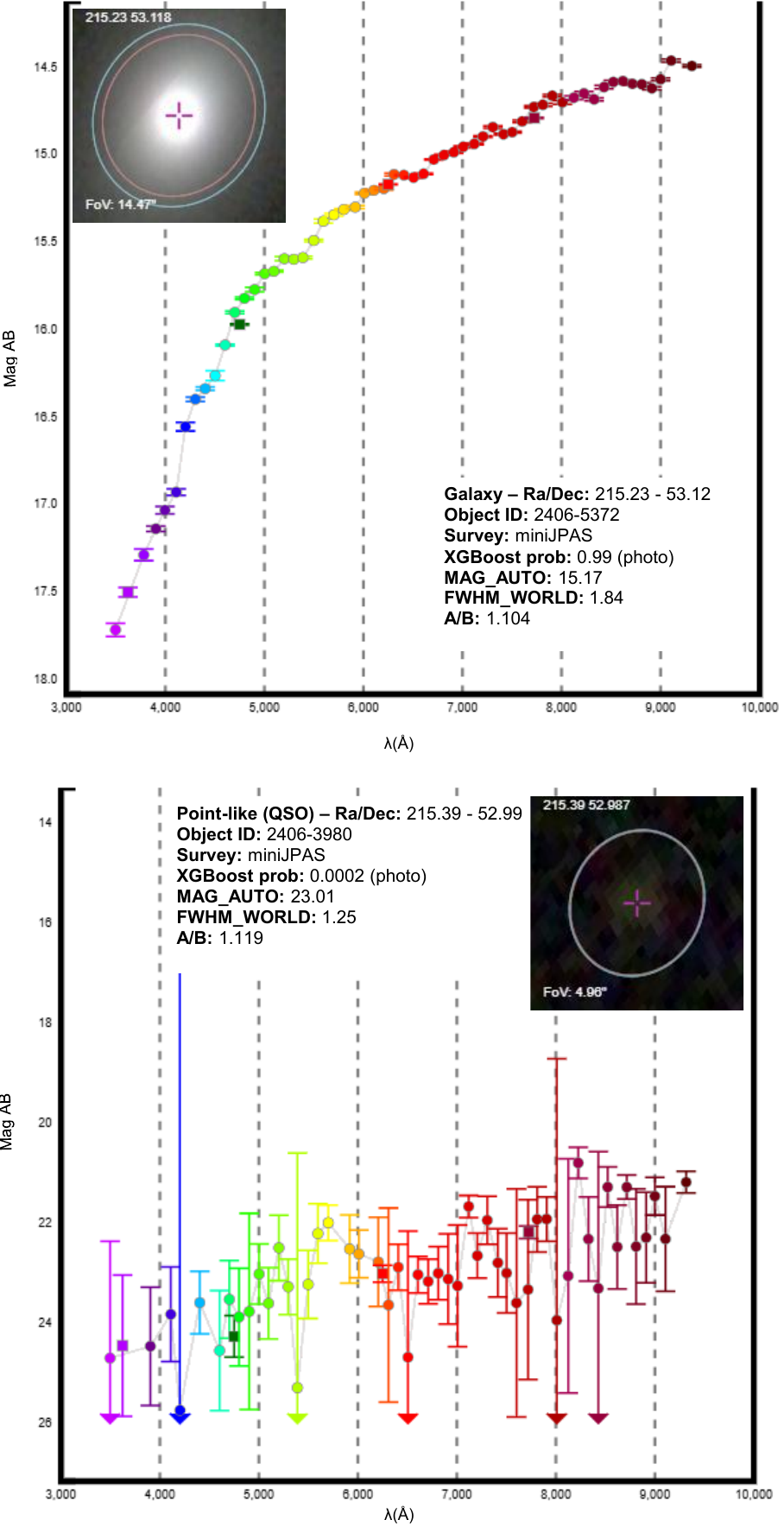}
\includegraphics[trim={0 0 0 0}, clip, height=11cm, width= .49\linewidth]{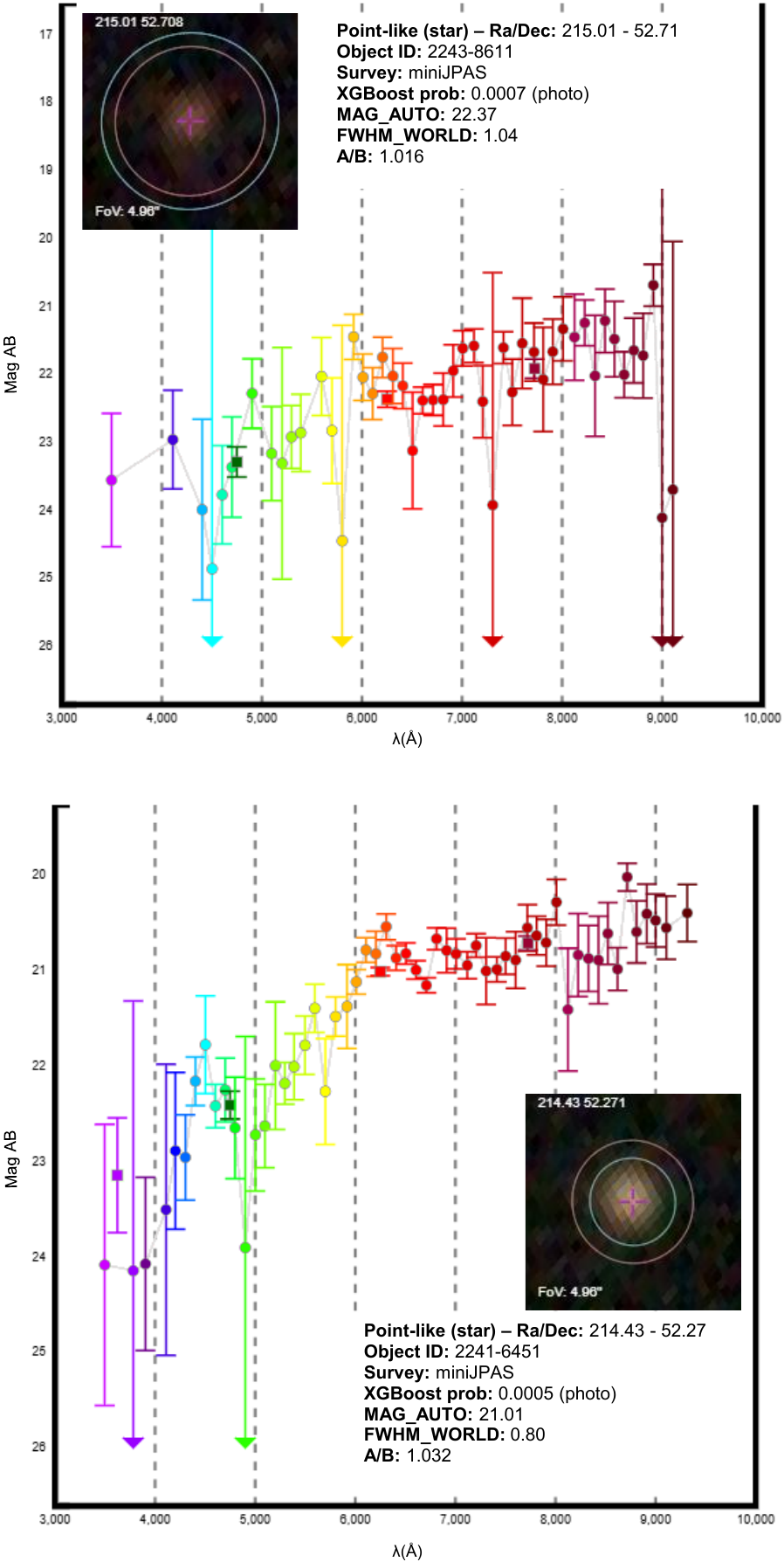}
\includegraphics[trim={0 0 0 0}, clip, height=5.3cm, width= .49\linewidth]{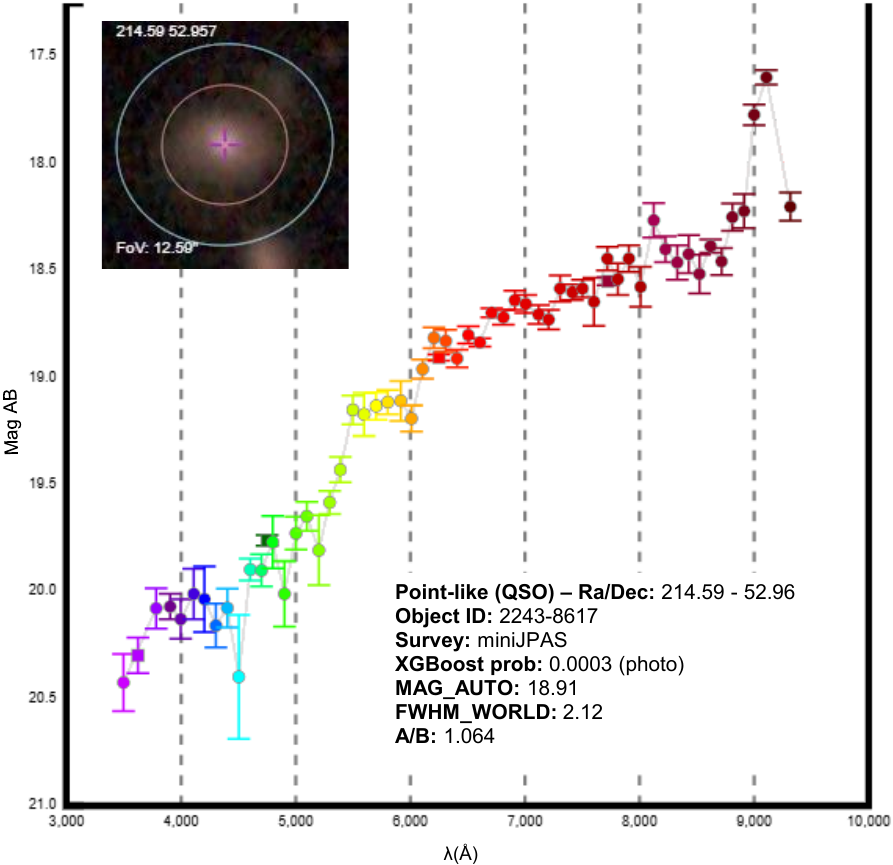}
\caption{
Representative examples of misclassified sources by the model trained with photometric features only. 
These cases illustrate typical limitations of a purely photometric approach.}
\label{fig:vismis-photo}
\end{figure}

\begin{figure}[p]
\raggedright 
\includegraphics[trim={0 0 0 0}, clip, height=11cm, width= .49\linewidth]{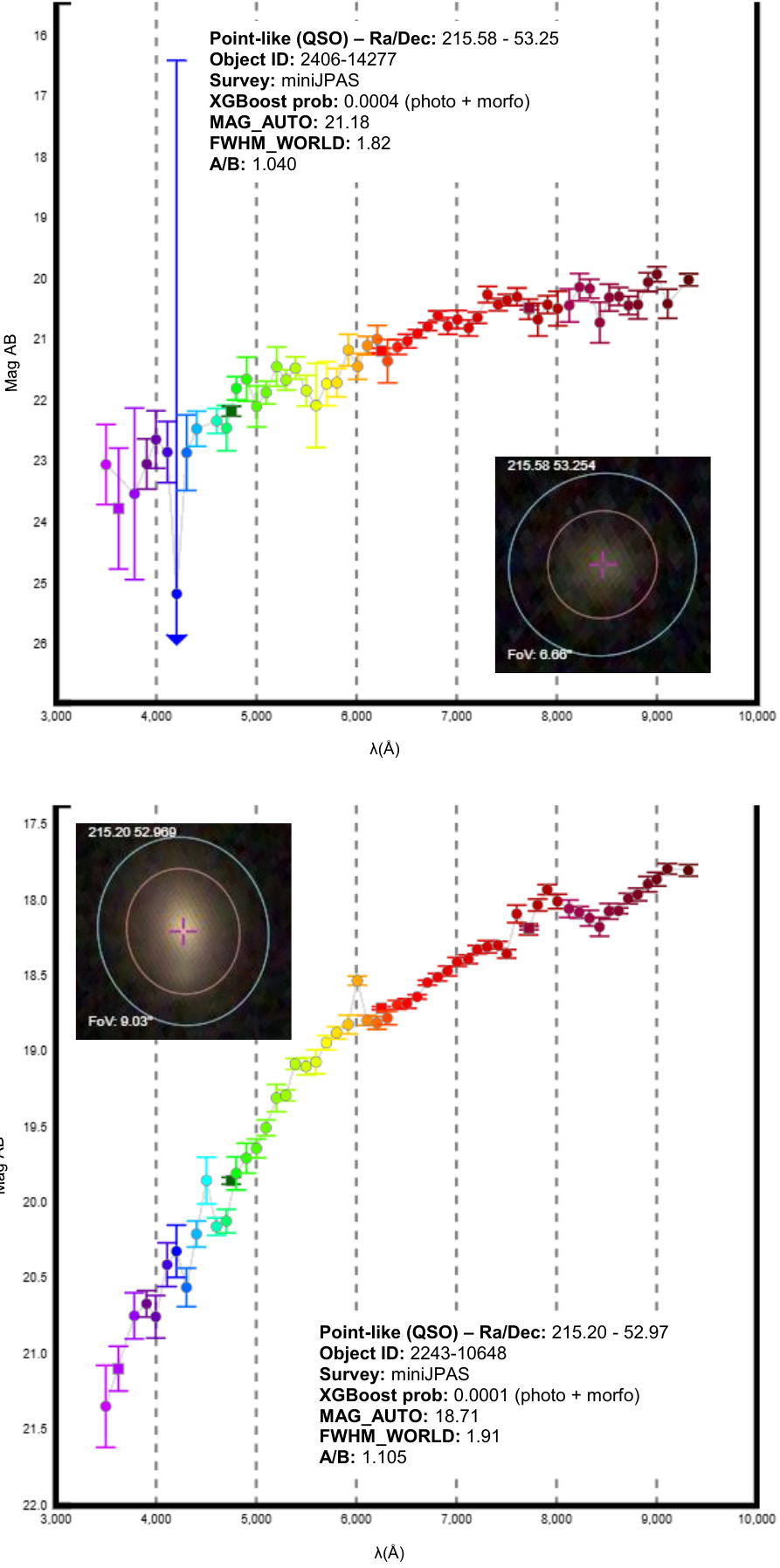}
\includegraphics[trim={0 0 0 0}, clip, height=11cm, width= .49\linewidth]{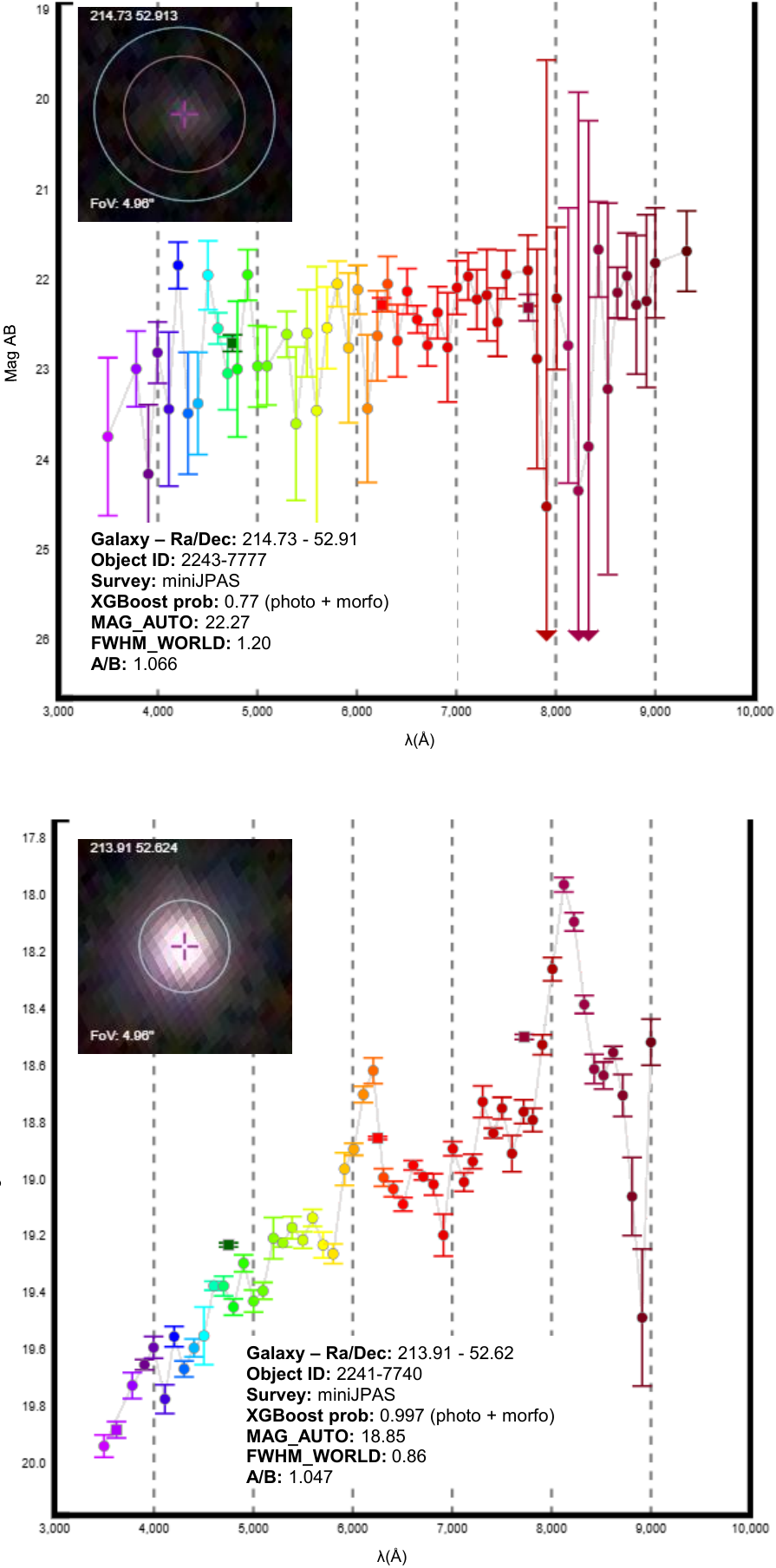}
\includegraphics[trim={0 0 0 0}, clip, height=11cm, width= .49\linewidth]{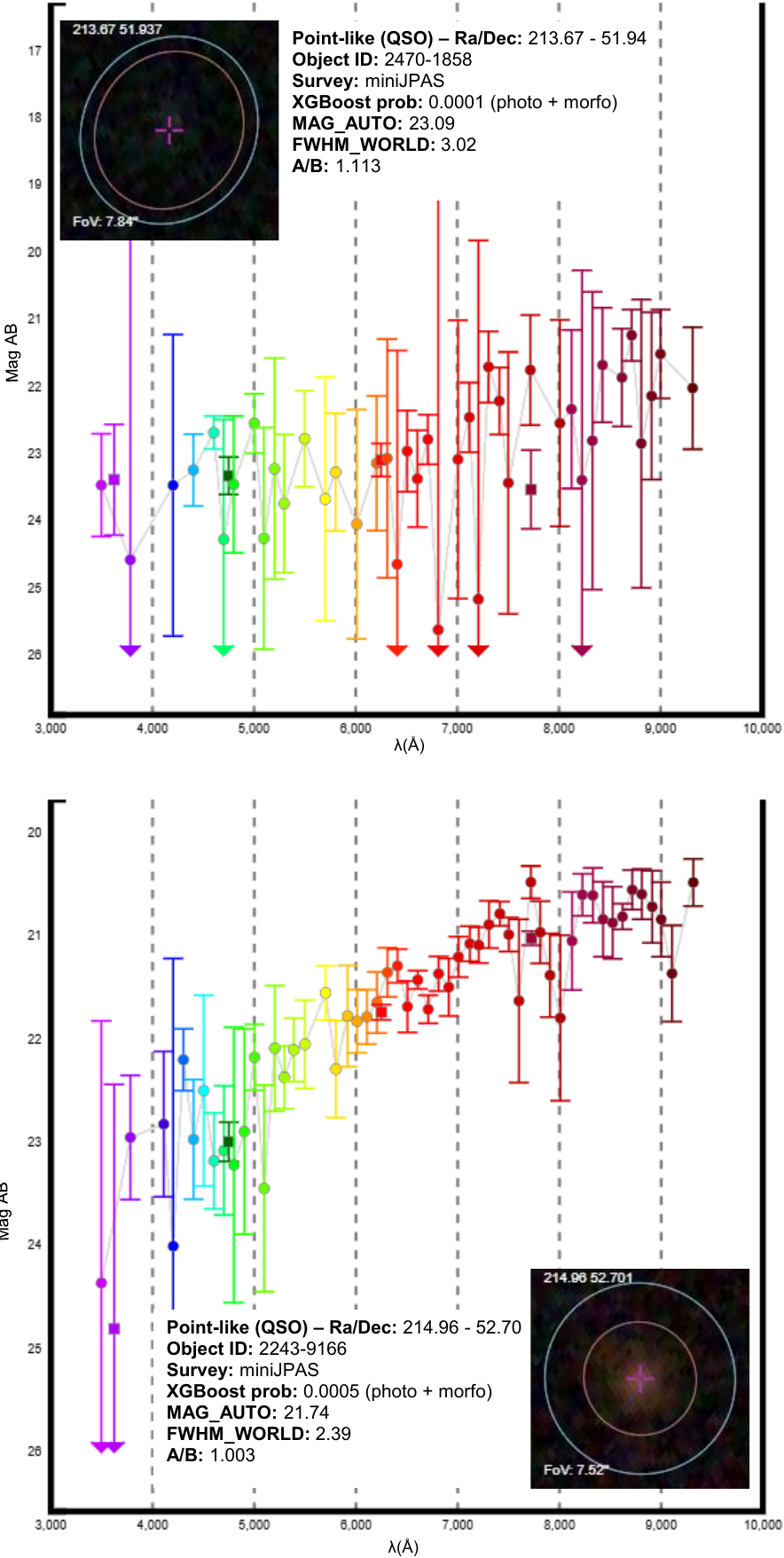}
\includegraphics[trim={0 0 0 0}, clip, height=5.3cm, width= .49\linewidth]{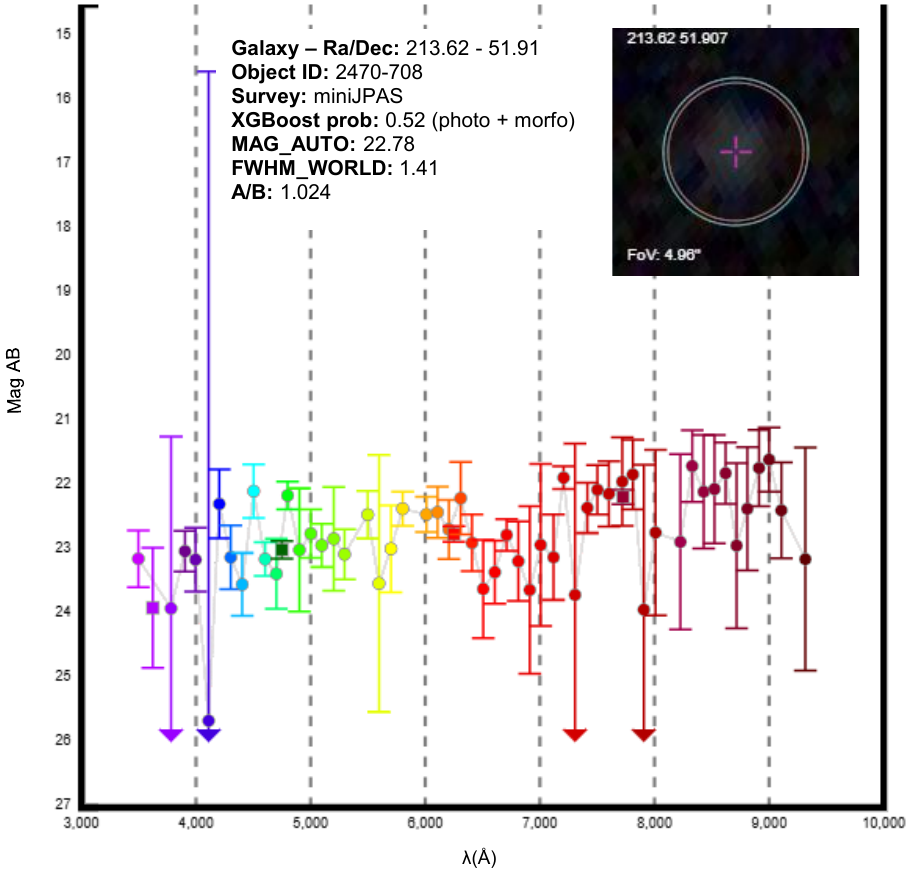}
\hfill
\caption{A Figure~\ref{fig:vismis-photo} but for the model trained with combined photometric and morphological features. These cases illustrate typical failure modes when morphology is included in the model.}

\label{fig:vismis-morpho}
\end{figure}

Each panel of Figures~\ref{fig:vismis-photo} and~\ref{fig:vismis-morpho} displays the spectral energy distribution across the 60 J-PAS bands, with broad-band points shown as squares, together with an RGB cutout from the $r$, $i$, and $g$ bands (inset). 
The examples include galaxies misclassified as point-like sources and point-like objects (stars or QSOs) misclassified as galaxies. 
For each case, we report the object ID, sky coordinates, survey origin, classifier probability (1 = star, 0 = galaxy), $r$-band magnitude, PSF, FWHM, and ellipticity $A/B$.

\begin{adjustwidth}{-\extralength}{0cm}

\reftitle{References}

\bibliography{refs}

\PublishersNote{}
\end{adjustwidth}
\end{document}